\documentclass[12pt,a4paper]{article}
\usepackage[dvips]{graphics}
\usepackage{cite}
\newcommand{\definmath}[2] {\def#1{\ifmmode#2\else$#2$\fi}}

\newcommand {\downto}
         {\mbox{ \begin{picture}(14,10)
                    \put(0,10){\line(0,-1){5.0}}
                    \put(2,5){\oval(4,4)[bl]}
                    \put(1,0){\makebox(0,0)[bl]{$\rightarrow$}}
                 \end{picture} }}

\mathchardef\ogon="012C%
\newbox\ogonbox
\setbox0\hbox{$\ogon$\kern-0.02em}
\setbox\ogonbox\hbox{\lower0.83\ht0\box0 }
\def\nonmathprzecinek#1{\ifx#1l{\ifnum \fam =\ttfam
\leavevmode \hbox {l\llap{-}}%
\else
\l%
\fi}%
\else
{\ifx#1L{\ifnum \fam =\ttfam
\leavevmode \hbox {L\llap{-\kern 0.1em}}%
\else
\L%
\fi}%
\else
{\leavevmode #1\llap {\copy\ogonbox}}%
\fi}%
\fi}%
\def\,{\ifmmode \let\next=\mathprzecinek
\else    \let\next=\nonmathprzecinek
\fi  \next}%
\begin{document}
\begin{titlepage}
\begin{center}
\mbox{\large EUROPEAN ORGANIZATION FOR NUCLEAR RESEARCH}
\end{center}
\begin{flushright}
CERN-EP/99-164\\
November 17, 1999\\
\end{flushright}
\vfill
\vfill
\begin{center}
{ \huge\bf
Leading Particle Production \\
\vspace*{5pt}
in Light Flavour Jets
}
\end{center}
\vfill
\begin{center}
{\bf\Large The OPAL Collaboration}
\end{center}
\vfill
\begin{center}
{\large\bf Abstract}\\
\end{center}
The energy distribution and type of the particle with the highest momentum in 
quark jets are determined for each of the five quark flavours making only 
minimal model assumptions.  The analysis is based on a large statistics sample 
of hadronic $\rm Z^0$ decays collected with the OPAL detector at the LEP 
$\rm e^+e^-$ collider.  These results provide a basis for future studies of 
light flavour production at other centre-of-mass energies.  We use our results 
to study the hadronisation mechanism in light flavour jets and compare the data 
to the QCD models JETSET and HERWIG.  Within the JETSET model we also directly 
determine the suppression of strange quarks to be
$$\rm \gamma _s  \ = \ 0.422 \pm 0.049 (stat.) \pm 0.059 (syst.)$$
by comparing the production of charged and neutral kaons in strange and 
non-strange light quark events.  Finally we study the features of baryon 
production.
\vfill
\vfill
\begin{center}
\bf (Submitted to Eur.\ Phys.\ Jour.\ C)
\end{center}
\vfill
\vfill
\end{titlepage}


\begin{center}{\Large        The OPAL Collaboration
}\end{center}\bigskip
\begin{center}{
G.\thinspace Abbiendi$^{  2}$,
K.\thinspace Ackerstaff$^{  8}$,
P.F.\thinspace Akesson$^{  3}$,
G.\thinspace Alexander$^{ 23}$,
J.\thinspace Allison$^{ 16}$,
K.J.\thinspace Anderson$^{  9}$,
S.\thinspace Arcelli$^{ 17}$,
S.\thinspace Asai$^{ 24}$,
S.F.\thinspace Ashby$^{  1}$,
D.\thinspace Axen$^{ 29}$
% ...
}\end{center}\bigskip
\bigskip
$^{  1}$School of Physics and Astronomy, University of Birmingham,
Birmingham B15 2TT, UK
\newline
$^{  2}$Dipartimento di Fisica dell' Universit\`a di Bologna and INFN,
I-40126 Bologna, Italy
\newline
$^{  3}$Physikalisches Institut, Universit\"at Bonn,
D-53115 Bonn, Germany
\newline
% ...
$^{  8}$CERN, European Organisation for Particle Physics,
CH-1211 Geneva 23, Switzerland
\newline
$^{  9}$Enrico Fermi Institute and Department of Physics,
University of Chicago, Chicago IL 60637, USA
\newline
% ...
$^{ 16}$Department of Physics, Schuster Laboratory, The University,
Manchester M13 9PL, UK
\newline
$^{ 17}$Department of Physics, University of Maryland,
College Park, MD 20742, USA
\newline
% ...
$^{ 23}$Department of Physics and Astronomy, Tel Aviv University,
Tel Aviv 69978, Israel
\newline
$^{ 24}$International Centre for Elementary Particle Physics and
Department of Physics, University of Tokyo, Tokyo 113-0033, and
Kobe University, Kobe 657-8501, Japan
\newline
% ...
$^{ 29}$University of British Columbia, Department of Physics,
Vancouver BC V6T 1Z1, Canada
\newline
% ...

\clearpage
\newpage
\section{Introduction}\label{sec:intro}

By identifying the flavour of the quark from which a jet develops one can 
experimentally test both electroweak and QCD theories. The power of flavour 
tagging has been demonstrated in many studies of bottom and charm quark 
production. Tagging light quark jets is experimentally much more difficult as 
these jets are not as distinctive as bottom quark or charm quark jets.  The 
main reason is that, unlike the heavy bottom and charm quarks, production of 
extra secondary up, down and strange quarks is abundant in jet development, 
making the identification of the hadron containing the primary quark ambiguous.
Due to these difficulties, tagging of individual light quarks has been studied 
and used in only a few analyses, for example 
in~\cite{bib-asOPAL,bib-ewOPAL,bib-afbDELPHI}. 

Whereas most of these analyses make assumptions about the details of 
hadronisation models, a method has been suggested in~\cite{bib-letmaett} 
which reduces the reliance on these assumptions.  This method has already been 
applied to determine the electroweak observables of individual light flavours 
by OPAL~\cite{bib-ewOPAL} at the $\rm e^+e^-$ collider LEP.  In the present 
analysis, high-energy $\pi ^{\pm}$, K$^{\pm }$, $\rm K^0_S$ mesons, protons 
and $\Lambda$ baryons are identified in the large $\rm Z^0$~data sample
and used as tagging particles.  In addition, high-momentum $\rm e^\pm$, 
$\mu^\pm$, $\rm D^{*\pm}$ mesons and identified bottom events are used to 
provide information about the heavy flavour backgrounds in these samples.  As 
suggested in~\cite{bib-field} and first confirmed by 
TASSO~\cite{bib-TASSOchgcor} and more precisely
studied in recent analyses, for example by SLD~\cite{bib-SLDchgcor},
these high-energy particles carry information about the original quark.  In 
this paper we extend the method used in Reference~\cite{bib-ewOPAL} to 
determine the probabilities $\eta ^i_q(x_p)$ for a quark flavour $q$ to 
develop into a jet in which the particle with the largest scaled momentum 
$x_p =2 p_i$/$\sqrt{s}$ is of type $i$.

The large number of $\rm Z^0$~decays collected at LEP and their well-known 
properties give a unique opportunity for determining the probabilities 
$\eta ^i_q(x_p)$.  From these measurements, we infer for the first time the 
flavour dependent fragmentation functions of light quarks.  This allows us to
study the hadronisation mechanism at an unprecedented level of detail.   
From these studies we determine in a direct way the suppression of strange 
quarks in the QCD sea and obtain insight into baryon production.

Apart from such hadronisation studies, the results may also be applied to
different environments.  After taking into account QCD scaling violations 
which can be rather precisely determined, the $\eta^i_q(x_p)$ from the 
$\rm Z^0$~allow one to calculate the $\eta ^i_q(x_p)$ at other centre-of-mass 
energies.  Possible applications include studies of light flavour production 
rates at other centre-of-mass energies~\cite{bib-letmaett} and the decay 
properties of the W boson, top quark or, if discovered, the Higgs boson.  

Section~\ref{sec:method} contains a summary of the method.
Section~\ref{sec:detector} describes the relevant features of the OPAL 
detector.  The event selection and the tagging particle identification 
are described in Section~\ref{sec:selection}.
The determination of the $\eta ^i_q(x_p)$ is described in Section~\ref{sec:eff} 
and their systematic uncertainties in Section~\ref{sec:syst}.  The results are 
shown and used to determine some properties of hadronisation in 
Section~\ref{sec:hadro_study}.


\section{Method}\label{sec:method}

As detailed in~\cite{bib-letmaett,bib-ewOPAL} the $\eta ^i_q(x_p)$
 are determined by 
using tags in event hemispheres\footnote{In this analysis, we denote
hemispheres as representing quark jets, since we are interested in studying 
the evolution of primary quarks into different hadron types.}. Each event is 
separated into two hemispheres using the plane perpendicular to the thrust 
axis containing the interaction point.  Each hemisphere is searched for the 
highest momentum particle, labelled $i$, subject to a minimum $x_p$ requirement.
If there are a number $N_q$ of hemispheres which originate from a quark of 
type $q$ and a number $N_{q\to i}(x_p)$ of tagging particles $i$ with a 
scaled momentum of at least $x_p \equiv x_{\rm cut}$ in these hemispheres, 
then the probability to find a tagging particle $i$ with a scaled momentum 
of at least $x_p$ is:
$$ \eta ^i_q(x_p) \ = \ \frac{N_{q\to i}(x_p)}{N_q}.  $$
The determination of the true $\eta ^i_q(x_p)$ at the ``generator level'', 
i.e.\ corrected for detector efficiencies and misassignment of the several 
particle types, is the main experimental aim of this paper.  The particles 
considered are those which have a high probability to tag light flavours: 
$\pi^\pm$, K$^\pm$, $\rm K^0_S$ mesons, protons, and $\Lambda$ baryons.  
Charge conjugation is implied throughout this paper.  What can be measured at 
the ``detector level'', i.e.\ before corrections for detector efficiencies
etc., are the number of hemispheres tagged by a particle of type $i$, labelled 
$N_i$ and called ``single-tagged hemispheres'', and the number of events 
containing a tagging particle in both hemispheres, labelled $N_{i j}$ and 
called ``double-tagged events'', where $i$ and $j$ are the tagging particle 
types.

These numbers are related to the probabilities:
\begin{eqnarray}
\label{eq:1}
{N_i \over N_{\rm had} }(x_p) & \ = \ &
2 \hspace*{-2mm} \sum_{q={\rm d,u,s,c,b}} 
\eta ^{i{\rm{,\ exp}}}_q(x_p) \thinspace  R_q \\
{\rm{and \ \ \ \ }} {N_{i j} \over N_{\rm had} }(x_p) & \ = \  &
(2-\delta_{ij})
\sum_{q={\rm d,u,s,c,b}} \rho_{ij}(x_p) \thinspace 
\eta ^{i{\rm{,\ exp}}}_q(x_p)  \thinspace 
\eta ^{j{\rm{,\ exp}}}_q(x_p)
\label{eq:2}
\thinspace  R_q ,
\end{eqnarray}
where $\delta_{ij}=1$ if $i=j$ and zero otherwise and $N_{\rm had}$ is the
number of hadronic $\rm Z^0$ decays.  The superscript `exp' denotes that the 
$\eta ^i_q(x_p)$
include possible distortions due to detector effects.  The parameters 
$\rho_{ij}(x_p)$ take into account correlations between the tagging
probabilities in opposite hemispheres, due to kinematic or geometrical effects, 
for example, and will not be equal to unity if such correlations exist.  
$R_q$ is the hadronic branching fraction of the $\rm{Z}^0$ to quarks $q$:
$$ R_q \ = \ \frac{\Gamma_{{\rm Z}^0\rightarrow q\bar q}}{\Gamma_{\rm had}}. $$
$R_c$ and $R_b$ are fixed to the LEP average measurements~\cite{bib-PDG}.
Given the good agreement of the Standard Model with data~\cite{bib-ewreview}, 
in particular the agreement of the measured $R_q$, we fix 
$R_{\rm d}/R_{\rm light}$, $R_{\rm u}/R_{\rm light}$ and 
$R_{\rm s}/R_{\rm light}$ to their predicted values~\cite{bib-ZFITTER}, 
such that $\sum_q R_q =1$, where 
$R_{\rm light}=R_{\rm d}+R_{\rm u}+R_{\rm s}$.

The true $\eta_q^i(x_p)$ are found after correcting for detector efficiencies
and misassignment of the tagged samples.  The relationship between the true 
$\eta ^i_q(x_p)$ and the observed $\eta ^{j{\rm{,\ exp}}}_q(x_p)$ is 
parametrised by a flow matrix, ${\cal E}^i_j$, which is taken from the 
simulation:
\begin{eqnarray}
\label{eq:flow}
\eta ^{j{\rm{,\ exp}}}_q(x_p)
 &=& \sum_i {\cal E}^i_j \eta ^i_q(x_p) \\
{\cal E}^i_j &=&  {N_{q\to i \to j}(x_p)^{\rm MC} \over 
                   N_{q\to i}      (x_p)^{\rm MC} },
\end{eqnarray}
where the sum over $i$ includes all tagging particle types at the generator 
level and $N_{q \to i\to j}(x_p)^{\rm MC}$ is the number of $q$-flavour 
Monte Carlo hemispheres tagged by particle $i$ at the generator level 
but $j$ in the detector.  ${\cal E}^i_j$ is found to vary slowly with $x_p$.  
In addition it is necessary to count events which are untagged at the generator 
level but still give rise to a tagging particle in the detector and will 
henceforth be denoted ``other background''.  For example, these events 
can be tagged by a particle which is not considered in this analysis, such as 
a stable hyperon (e.g.\ $\Sigma^-$, $\Xi^-$), or are tagged by a high-momentum
particle which has a true momentum slightly below the minimum required
$x_p$ due to the finite momentum resolution of the detector.  

The system of equations (\ref{eq:1}) and (\ref{eq:2}) has 20 equations 
(5 single and 15 double tags) with 25 unknown $\eta ^i_q(x_p)$ for the five 
quark flavours produced in $\rm Z^0$~decays.  We extend the system of equations 
in the following two ways:
\begin{enumerate}

\item In order to better measure the heavy flavour $\eta_q^i(x_p)$, we include 
charm and bottom tags by identifying $\rm{D}^{*\pm}$ mesons, or a vertex 
displaced from the interaction point.  These techniques have been used 
previously in OPAL papers~\cite{bib-OPALD*,bib-OPALRb} and are briefly 
described in Section~\ref{sec:cbtag}.  Charged leptons $\rm e^\pm$ and 
$\mu^\pm$, which mainly tag heavy flavours but are still a source of
background in the light flavour charged hadron samples, are also identified 
and included in the equation system.  Note that the vertex tag does
not depend on $x_p$.

\item In order to reduce the number of unknown $\eta_q^i(x_p)$, we invoke 
hadronisation symmetries such as 
$\eta_{\rm{d}}^{\pi^\pm}=\eta_{\rm{u}}^{\pi^\pm}$,
which are motivated by the flavour independence of QCD and SU(2) isospin 
symmetries.  They have been extensively discussed in~\cite{bib-letmaett} for 
$x_p>0.5$.  At lower momenta the relations are potentially broken by isospin 
violating decays, for example $\phi(1020)$ to charged and neutral kaons.
Nevertheless, the relations 
\begin{eqnarray*}
\eta_{\rm{d}}^{\pi^\pm}   &=& \eta_{\rm{u}}^{\pi^\pm}, \\
\eta_{\rm{s}}^{\rm K^\pm} &=& \eta_{\rm{s}}^{\rm K^0} {\rm \ and} \\
\eta_{\rm{d}}^{\rm e^\pm} &=& \eta_{\rm{u}}^{\rm e^\pm}
\end{eqnarray*}
are expected to be valid to high precision also after decays.  This has been 
checked using the QCD model JETSET~\cite{bib-JETSET} after adjusting the yield 
and energy dependence of prominent resonances to the measurements at 
LEP~\cite{bib-hadatLEP}.  We find the relations hold to within 2\% above 
$x_p=0.2$, the range used in this analysis. Here $\rm K^0$ is made up of both 
$\rm K^0_S$ and $\rm K^0_L$, which are assumed to be equal.  A relation that is
used which is violated by up to 10\% at low $x_p\approx 0.2$ due to decays is
$$\eta_{\rm{d}}^{\Lambda(\overline{\Lambda})} = 
  \eta_{\rm{u}}^{\Lambda(\overline{\Lambda})}.$$
When introducing these hadronisation symmetries into the equation system,
we make whatever small corrections for isospin-violating decays are necessary
according to the JETSET Monte Carlo.  The HERWIG model~\cite{bib-HERWIG} is not 
used to check the hadronisation symmetries because it violates SU(2) isospin 
symmetry for technical reasons~\cite{bib-LEP1yellow}.

\end{enumerate}

These additions give a total of 54 equations with 41 unknown $\eta^i_q(x_p)$. 
The equations are solved requiring a minimum $x_p > 0.2$, 0.3, 0.4, 0.5 and 
0.6.  In this paper, full results are presented for a minimum $x_p > 0.2$ and 
summarised for the other cut values.  

 
\section{The OPAL Detector}\label{sec:detector}
 
The OPAL detector is described in detail in~\cite{bib-OPALDET}.  The relevant 
features for this analysis are summarised in this Section.  OPAL uses a 
right-handed coordinate system, where the $z$-axis points along the electron 
beam, $r$ is the coordinate normal to this axis, and $\theta$ and $\phi$ are 
the polar and azimuthal angles with respect to $z$.

The central tracking system, inside a 0.435\thinspace{T} axial magnetic field, 
provides 
a charged track momentum resolution of $\sigma_p/p=0.02 \oplus 0.0015~p_t$,  
where $p_t$ is the momentum component perpendicular to the beam axis in GeV.
A silicon microvertex detector~\cite{bib-OPALSI}, close to the interaction 
point, is surrounded by three drift chambers: a vertex detector, a large volume 
jet chamber which provides up to 159 space points per track, and $z$-chambers 
which give a precise measurement of the polar angle of charged tracks.  The 
large number of samplings in the jet chamber also provides a determination 
of the specific ionisation energy loss, d$E$/d$x$, with a resolution of
$\sigma({\rm{d}}E /{\rm{d}}x)/({\rm{d}}E /{\rm{d}}x) \sim
0.032$~\cite{bib-OPALDEDX} in multihadronic events for tracks with 
$|\cos\theta|<0.7$ and the maximum number of samplings. At larger 
$|\cos\theta|$ the resolution is degraded because fewer measured points 
are available.  The d$E$/d$x$ measurements have been calibrated using
almost pure control samples of, for example, pions from $\rm K^0_S$, 
$\mu$-pair events  and photon conversions into electrons, such that the central 
values are known to $0.10\sigma({\rm{d}}E/{\rm{d}}x)$ and the resolution 
to a precision of 10$\thinspace\%$~\cite{bib-TN562}.  The electromagnetic 
calorimeter consists of 11\,704 lead glass blocks, each subtending 
a solid angle of $40\times40$\thinspace mrad$^2$. The muon chambers surround 
the calorimeter, behind approximately eight absorption lengths of material.

Detector efficiencies and possible detector biases are studied with
approximately six million simulated hadronic $\rm Z^0$ decays generated with 
the JETSET 7.4 model~\cite{bib-JETSET} and passed through a detailed simulation
of the OPAL detector~\cite{bib-GOPAL}. The fragmentation parameters have been 
tuned to describe event shapes and other distributions as described 
in~\cite{bib-toon}.  In addition, for fragmentation studies one
million fully-simulated hadronic $\rm Z^0$ decays generated with the 
HERWIG 5.8~\cite{bib-HERWIG} Monte Carlo generator are used.


\section{Event Selection and Tagging Methods}\label{sec:selection}

The analysis uses approximately 4.1 million multihadronic Z$^0$ 
decays collected between 1991 and 1995.  The standard OPAL multihadronic 
selection is applied~\cite{bib-MHSELECT}.  To select events which are well 
contained in the detector, the polar angle of the thrust axis, 
$\theta_{\rm T}$, calculated using charged tracks and electromagnetic 
calorimeter clusters which have no associated track in the jet chamber is 
required to satisfy $|\cos\theta_{\rm T}|<0.8$.  To assure good bottom quark 
tagging quality it was also required that the silicon microvertex detector be
functioning well. 

In this analysis we select tagging particles with $x_p>0.2$.  The 
selection of these highly energetic particles enhances the background fraction 
from $\rm{Z}^0\to\tau^+\tau^-$ events, so we require in addition each event to 
have at least eight well-measured tracks~\cite{bib-goodtracks}. With these 
requirements, 2\thinspace 820\thinspace 220 events are retained.  In this event 
sample the $\tau$ background is reduced to less than 0.03\thinspace\%, as 
estimated using fully simulated events generated with the KORALZ Monte Carlo 
generator~\cite{bib-KORALZ}, and so can be neglected.

Next a high-energy stable particle or a charm or bottom tag is required. 
The selection of particles was optimised for the highest accuracy of the 
desired $\eta ^i_q(x_p)$, balancing the potential loss in separation power 
against efficiencies.  As discussed in Section~\ref{sec:method}, we look for 
the particle ($\pi^\pm$, $\rm K^\pm$, $\rm p(\bar{p})$, $\rm K^0_S$, 
$\Lambda(\bar{\Lambda})$, $\rm e^\pm$, $\rm \mu^\pm$ or $\rm D^{*\pm}$, )
in each event hemisphere with the highest scaled momentum $x_p$.  
To ensure good charged pion, kaon and proton separation, and reliable 
$\rm K^0_S$ and $\Lambda$ reconstruction, we require that the tagging 
particles have polar angles $|\cos\theta|<0.9$.  

\subsection{Stable hadrons}\label{sec:particleid}

The d$E$/d$x$ measurement of good quality tracks is used to identify charged
pions, charged kaons and protons.  For each track the d$E$/d$x$ weight $w_i$ 
is used to separate the particle types. The weight is the $\chi^2$ probability 
that the track is consistent with a hypothesised particle $i$. 
We require:
\begin{itemize}
\item for pion candidates:
               $w_{\pi^{\pm}}>0.1$ and
               $w_{\rm K^{\pm}}<0.1$;
\item for kaon candidates:
               $w_{\rm K^{\pm}}>0.1$ and 
               $w_{\pi ^{\pm}}<0.1$;
\item for proton candidates:
               $w_{\rm p( \overline{\rm p})}>0.1$ and
               $w_{\rm K^{\pm }}<0.1$.
\end{itemize}
These selection criteria give three disjoint samples.  

Averaged over all five quark flavours,
$$
{\cal P}^i_j(x_p) = { \sum_q {\cal E}^i_j  \eta^i_q(x_p) \over
                      \sum_q  \eta_q^{j,\ \rm exp}(x_p)}
$$
express the probability that a particle identified as type $j$ stems from a 
true particle type $i$.  The values are given in Table~\ref{tab:flowmtx02} 
for samples with $x_{p}>0.2$.  The determination of the flow matrix
${\cal E}^i_j$, which is taken from simulation, was discussed in 
Section~\ref{sec:method}.  As also 
discussed in Section~\ref{sec:method}, a few percent of the tagging particles 
have a true $x_p$ value below the cut imposed on the measured $x_p$ value but 
are tagged in the detector due to the finite momentum resolution.  The Monte 
Carlo also predicts that there is a 1\% contamination from charged hyperons, 
mostly $\Sigma^-$, in the proton sample.  These sources of background are 
included in ``other background'' given in Table~\ref{tab:flowmtx02}.  The cuts, 
including the event and thrust axis cut, lead to the efficiencies shown in the 
bottom row of Table~\ref{tab:flowmtx02}, which are defined as the number of 
hemispheres which are correctly tagged at the detector level divided by the 
number of hemispheres which are tagged at the generator level.

\subsection{Electron and muon identification}

Electrons are identified using a number of discriminating variables, 
principally the d$E$/d$x$ and the energy loss in the electromagnetic 
calorimeter~\cite{bib-OPALRb}.  Muons are selected by matching tracks in
the central detector with hits in the muon chambers~\cite{bib-OPALRb}.
As can be seen from Table~\ref{tab:flowmtx02}, the efficiencies to correctly 
tag an electron or a muon are about 20\% and 70\% in this hadronic jet
environment, respectively, with purities of around 60$\%$. 

\subsection{\boldmath $\rm K^0_S$ and $\Lambda$ identification}

The procedures to identify the weakly decaying $\rm K^0_S$ and $\Lambda$ are 
described in~\cite{bib-OPALK0S} and~\cite{bib-OPALLAMBDA}, respectively. The 
decays $\rm K^0_S\to\pi^+\pi^-$ and $\Lambda\to\rm{p}\pi^-$ are reconstructed
by combining two oppositely-charged tracks  which have a crossing point in the 
plane orthogonal to the beam axis.  If a secondary vertex is found, the 
invariant masses $m_{\pi^+\pi^-}$ and $m_{ {\rm{p}}\pi^-}$ of the $\pi^+\pi^-$ 
and $ {\rm{p}}\pi^-$ mass assignments are calculated.  $\rm K^0_S$ 
candidates are required to have invariant masses in the ranges 430\thinspace 
MeV~$<m_{\pi^+\pi^-} <$~570 \thinspace MeV and 
$m_{ {\rm{p}}\pi^-} > 1.13$\thinspace GeV, to reduce the contamination from 
$\Lambda\to{\rm{p}}\pi^-$ decays. Similarly, all candidates which have
1.10~GeV~$<m_{ {\rm{p}}\pi^-}<$~1.13\thinspace GeV are accepted as $\Lambda$ 
candidates.   The $\rm K^0_S$ selection in~\cite{bib-OPALK0S} is extended in
the present analysis to $|\cos\theta|<0.9$ from $0.7$, resulting in a slightly 
worse overall mass resolution, but the acceptance is increased and is the same 
as the other particle tags used, thus reducing geometric hemisphere 
correlations.  The combinatoric backgrounds are estimated from the Monte Carlo 
and cross-checked by determining the backgrounds from candidates with invariant 
masses in sidebands around the signal.

\subsection{Charm quark and bottom quark tags}\label{sec:cbtag}

The sample enriched in charm quark events is found by selecting hemispheres 
with a high-energy $\rm D^{*\pm}$~\cite{bib-OPALD*}.  The decay modes and 
the cuts used on the $x_p^{\rm D^{*}}$ values, which are calculated from the 
measured decay products of the $\rm D^{*\pm}$ candidate, are
\begin{center}
\begin{tabbing}
  \hspace{5cm} \= \hspace{5cm} \= \kill
  \> ${\rm D^{*{\scriptscriptstyle +}}} \rightarrow {\rm D^0}\pi^+$ \\
  \> $\phantom{{\rm D^{*{\scriptscriptstyle +}}} \rightarrow }\hspace{4pt} 
            \downto {\rm K^-}\pi^+$\                     \> $x_p^{\rm D^{*}} > 0.4 $  \\
  \> $\phantom{{\rm D^{*{\scriptscriptstyle +}}} \rightarrow }\hspace{4pt} 
            \downto {\rm K^-}{\rm e}^+ \nu_{{\rm e}}$\   \> $x_p^{\rm D^{*}} > 0.4 $\\
  \> $\phantom{{\rm D^{*{\scriptscriptstyle +}}} \rightarrow }\hspace{4pt} 
            \downto {\rm K^-}\mu^+\nu_{\mu}$\            \> $x_p^{\rm D^{*}} > 0.4 $ \\ 
  \> $\phantom{{\rm D^{*{\scriptscriptstyle +}}} \rightarrow }\hspace{4pt} 
            \downto {\rm K^-}\pi^+$\                     \> $x_p^{\rm D^{*}} > 0.4 $  \\
  \> $\phantom{{\rm D^{*{\scriptscriptstyle +}}} \rightarrow }\hspace{4pt} 
            \downto {\rm K^-}\pi^+\pi^-\pi^+$\           \> $x_p^{\rm D^{*}} > 0.5 $
\end{tabbing}
\end{center} 
In the simulation the $\rm D^{*\pm}$ tagging efficiency is about 1\% for a 
charm quark purity of about 58\%.  

The bottom quark tag uses a number of discriminating variables calculated 
from a reconstructed secondary vertex~\cite{bib-OPALRb}.  From the number of 
double and single-tagged events the hemisphere tagging efficiency is found 
to be about 19\% for a bottom jet purity of about 96\%.  These efficiencies 
and purities are used only for the cross-check outlined in 
Section~\ref{ssec:claudio}.  The hemispheres tagged as bottom are counted 
even if they are already present in the high $x_p$ tagged samples.


\section{Determination of \boldmath $\eta^i_q(x_p)$}\label{sec:eff}

The numbers $N_i$ and $N_{ij}$ of measured single- and double-tagged events are 
given in Table~\ref{tab:evcount_x02} for $x_p>0.2$.  They are used as input to 
the equation system which is solved for the $\eta^i_q$ by using a $\chi^2$ fit. 
The $\chi^2$ function is defined as:
\begin{eqnarray}
\chi^2 &=&
	  \sum_i \Bigl[ {\tilde{N}_i - 2 N_{\rm had} \sum_q R_q \thinspace 
	  \tilde{\eta}^{i{\rm{,\ exp}}}_q(x_p)
        \over \sqrt{\tilde{N}_i}} \Bigr] ^2 \\
&+&
	  \sum_{i,j} \Bigl[ {N_{ij} - (2-\delta _{ij}) N_{\rm had} 
	  \rho_{ij}(x_p) \sum_q \thinspace R_q \thinspace 
	  \eta ^{i{\rm{,\ exp}}}_q(x_p) \thinspace 
	  \eta ^{j{\rm{,\ exp}}}_q(x_p) \over \sqrt{N_{ij}}} \Bigr] ^2 
\label{eq:chi2}
\end{eqnarray}
where
\begin{eqnarray*}
\tilde{N}_i &=& N_i - \sum_j (1+\delta_{ij}) N_{ij} \ \ \ \ {\rm and}\\
\tilde{\eta}^{i{\rm{,\ exp}}}_q(x_p)
	  &=& \eta ^{i{\rm{,\ exp}}}_q(x_p) - \sum_j \rho_{ij}(x_p) 
	  \thinspace \eta ^{i{\rm{,\ exp}}}_q(x_p)
	  \thinspace \eta ^{j{\rm{,\ exp}}}_q(x_p)
\end{eqnarray*}
are used to correct for double-counting of hemispheres in the single- and
double-tagged samples.  These two corrections are necessary to remove 
double-tagged events from the sample of single-tagged hemispheres.  

In addition, the hadronisation symmetries given in Section~\ref{sec:method} are
used after being corrected for detector effects and making small corrections 
for isospin-violating decays according to the JETSET Monte Carlo.  Furthermore, 
certain very small, and therefore unmeasurable, $\eta ^i_q$ are fixed to their 
JETSET values at the generator level, namely: $\eta_{\rm d,u,s}^{\mu^\pm}$,
$\eta_{\rm d,u,s}^{\rm{D}^{*\pm}}$ and $\eta_{\rm d,u,s}^{\rm b-vtx}$.  
The $\rho_{ij}$ parameters, parametrising possible kinematic and geometrical 
correlations, are taken from the simulation.  Geometrical correlations
lead in general to a positive correlation $\rho_{ij}\ge 1$.  Motivated by 
simulation studies, the correlation is assumed to be the same for all tagging 
particle types except for the $\rm D^{*\pm}$ and the bottom tag.   
Typical values are $\rho_{ij}=1.020\pm 0.002$ at $x_p>0.2$ and 
$\rho_{ij}=1.13 \pm 0.03 $ at $x_p>0.5$, where the errors are from Monte 
Carlo statistics, and $i$ and $j$ run over all tagging particle types except 
for $\rm D^{*\pm}$ and the bottom vertex tag.  For the $\rm D^{*\pm}$ and 
bottom vertex tags, the correlations are determined individually for each 
measured double-tagged sample.  For example, $\rho_{\pi^\pm {\rm{D}^{*\pm}}} 
= 1.048 \pm 0.021$ and $\rho_{\pi^\pm {\rm b-vtx}} = 1.018 \pm 0.006$ for 
$x_p^\pi>0.2$, where the errors are again from Monte Carlo statistics.  The 
extracted $\eta ^{i{\rm{,\ exp}}}_q(x_p)$ are corrected for the detector 
effects using the flow matrix ${{\cal E}}^i_j$ in equation~\ref{eq:flow}.  

The results after corrections for detector efficiency and 
particle misassignment are listed in Table~\ref{tab:results02} for $x_p>0.2$.
The table also includes the statistical and systematic uncertainties and a 
comparison with the JETSET and HERWIG models.  We give details of the results 
only for the tagging particle types which mainly tag light flavours, namely 
$\pi^\pm$, $\rm K^\pm$, $\rm K^0_S$, proton and $\Lambda$.  The statistical 
correlations between the parameters are given in Table~\ref{tab:corr02}.  

\begin{figure}[p]
\begin{center}
\resizebox{\textwidth}{!}
{\includegraphics{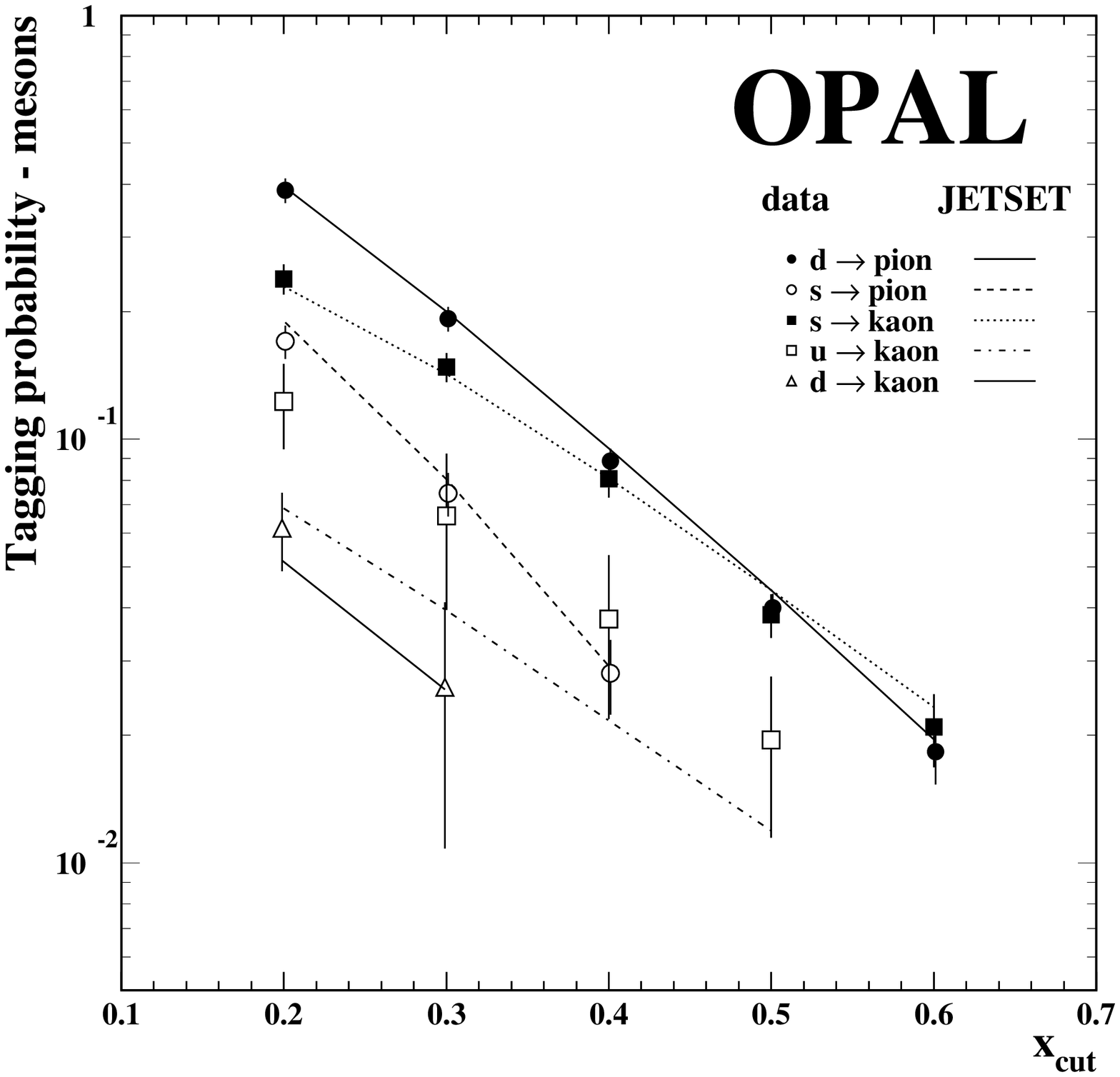}}
\caption{Tagging probabilities as a function of 
the minimum $x_p$ cut for charged pions and kaons.
Data points are correlated for different values of the minimum $x_p$ cut.
The errors shown are statistical plus systematic.
The lines show the JETSET predictions.
\label{fig:tag-eff-int-mes1} }
\end{center}
\end{figure}

\begin{figure}[p]
\begin{center}
\resizebox{\textwidth}{!}
{\includegraphics{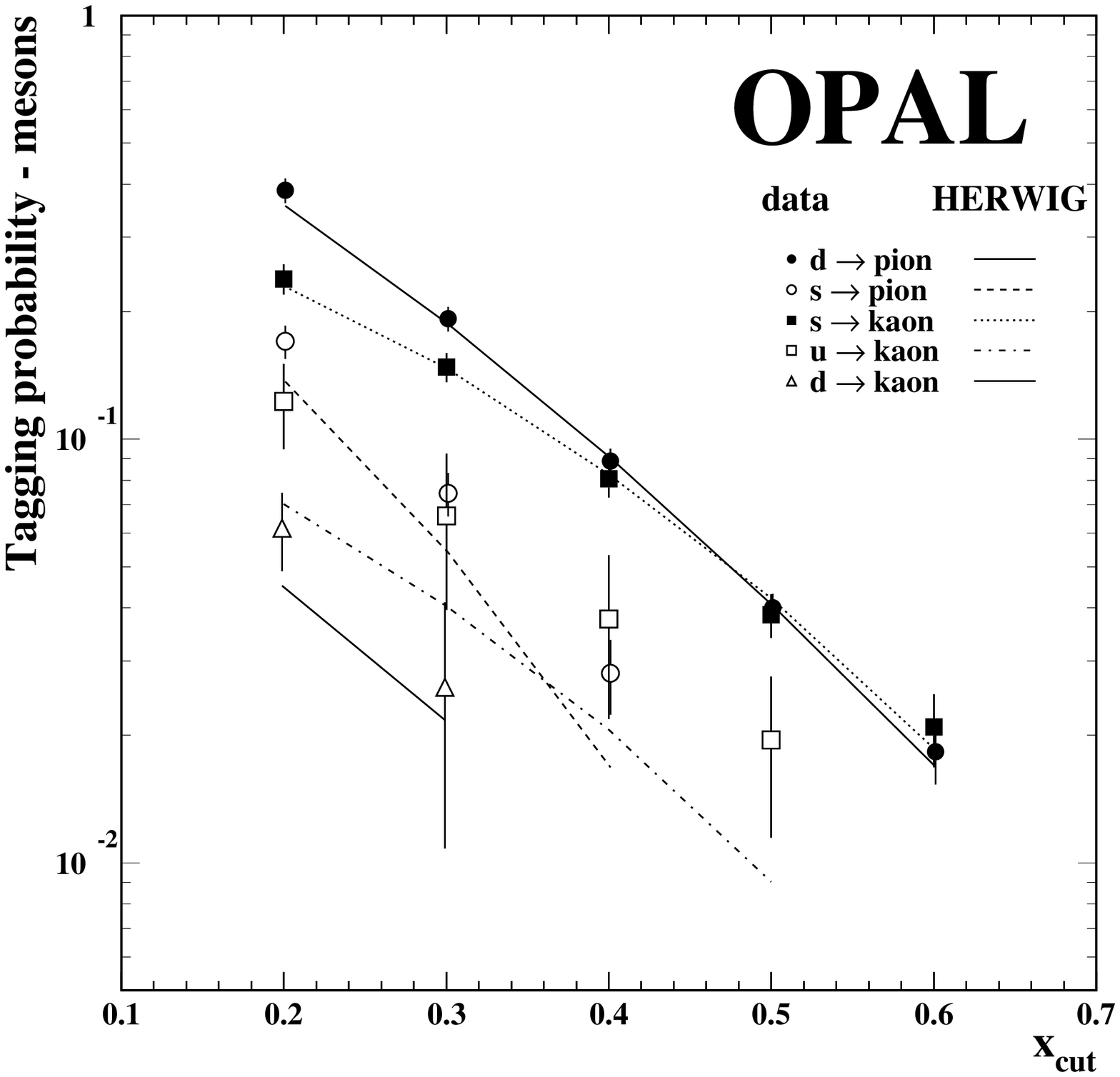}}
\caption{Tagging probabilities as a function of 
the minimum $x_p$ cut for charged pions and kaons.
Data points are correlated for different values of the minimum $x_p$ cut.
The errors shown are statistical plus systematic.
The lines show the HERWIG predictions.
\label{fig:tag-eff-int-mes2} }
\end{center}
\end{figure}

\begin{figure}[p]
\begin{center}
\resizebox{\textwidth}{!}
{\includegraphics{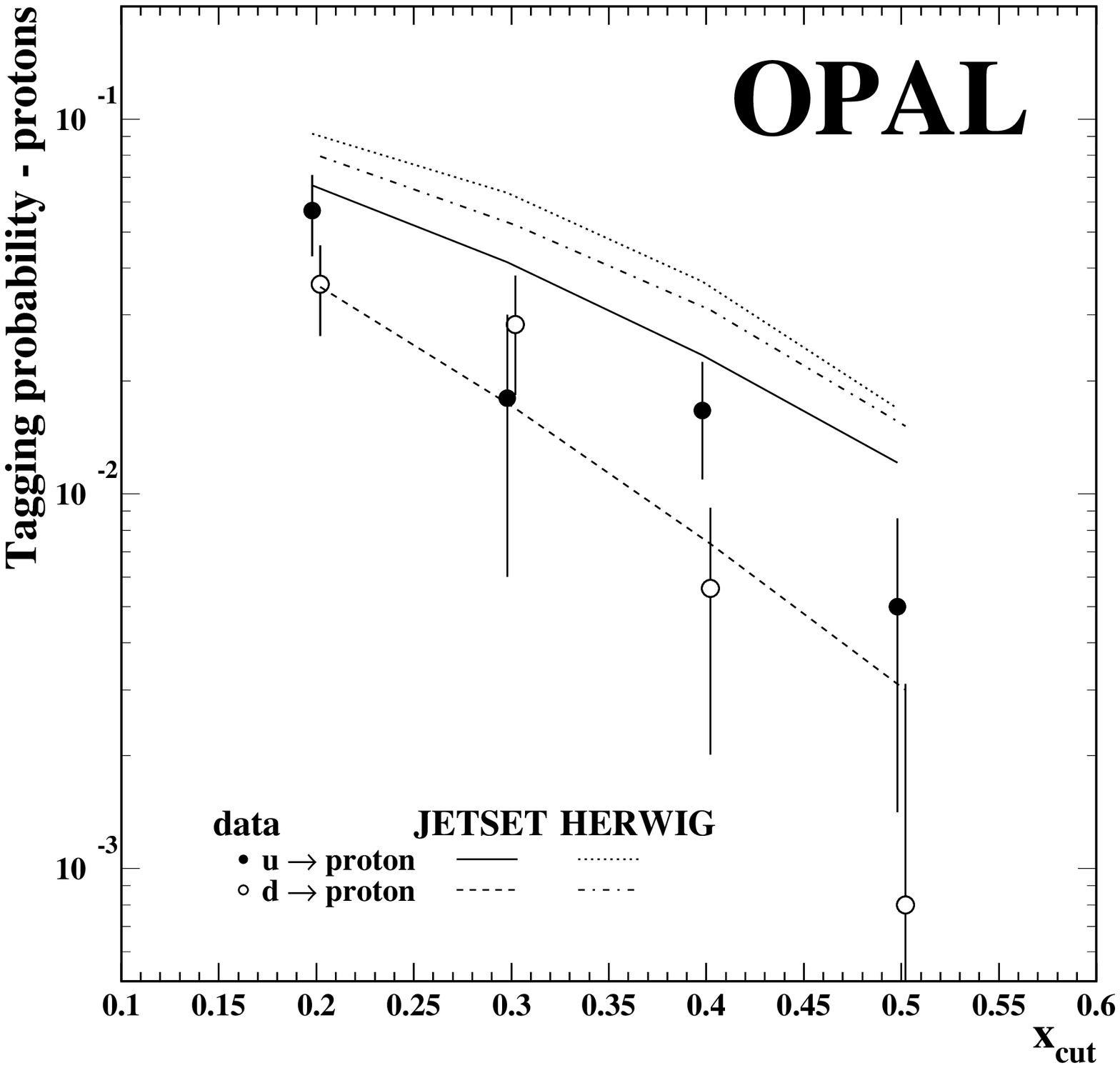}}
\caption{Tagging probabilities as a function of 
the minimum $x_p$ cut for protons. 
Data points are correlated for different values of the minimum $x_p$ cut.
The errors shown are statistical plus systematic.
The lines show the JETSET and HERWIG predictions.
\label{fig:baryon_int1} }
\end{center}
\end{figure}

The corrected results also for $x_p$ cuts other than $x_p>0.2$ are summarised 
in Table~\ref{tab:xintegral} with statistical and systematic error combined.  
Some of the larger $\eta ^i_q(x_p)$ are shown in 
Figures~\ref{fig:tag-eff-int-mes1}-~\ref{fig:baryon_int2}.  Correlations 
between the $\eta ^i_q(x_p)$ for different particle types and between the 
values obtained with different $x_p$ cuts are discussed in 
Section~\ref{sec:hadro_study}.

The solutions were checked to be unique and that the error matrices were
positive definite.  The $\chi^2$ per degree of freedom of the solutions 
are typically $\approx 1.2$, and are given in Table~\ref{tab:xintegral}.


\section{Systematic Uncertainties}\label{sec:syst}

The validity of the method used in this paper is tested using approximately 
six million hadronic $\rm{Z}^0$ decays generated using the JETSET Monte Carlo
and including a full simulation of the OPAL detector.  The $\eta^i_q(x_p)$ 
obtained from solving the equation system agree with the Monte Carlo 
predictions.

\subsection{Main uncertainties}

The main sources of systematic uncertainty are due to the limited 
knowledge of the efficiencies and purities of the particle identification.
Others are due to the flavour composition of the $\rm D^{*\pm}$
and bottom-tagged samples.  The third class of uncertainties is related 
to opposite-hemisphere correlations in the double-tagged samples.
Since these classes of errors are largely uncorrelated, we estimate
their individual impact on the $\eta ^i_q(x_p)$
and add them quadratically to obtain 
the overall systematic error.  The errors are determined by changing in turn 
each input parameter according to the estimated individual range of uncertainty,
repeating the analysis, and interpreting the shifts as the error contribution.  
A break down of the individual error contributions for the most important 
$\eta ^i_q(x_p)$ is listed in Table~\ref{tab:systerr02} for $x_{p}>0.2$.  
Relative contributions to the systematic error at other minimum values of 
$x_p$ are similar.

The following systematic uncertainties are considered:
\begin{itemize}

\item {\bf Charged particle purity and efficiency:}

Systematic errors are applied to the charged pion, charged kaon, and proton 
yields.  The uncertainties in these corrections are estimated by varying the 
widths and mean values of the ionisation energy loss in the simulation according
to the uncertainties discussed in Section~\ref{sec:detector}~\cite{bib-TN562}. 
These errors are the dominant ones for all $\eta ^i_q(x_p)$ of charged hadrons.

The uncertainties in the electron and muon identification have been discussed 
in~\cite{bib-OPALRb}.  The error of the electron identification efficiency is 
due primarily to uncertainties in modelling the d$E$/d$x$. The modelling of the 
muon efficiency has been checked using $\mu$-pair and 
$\rm e^+e^-\to e^+e^-\mu^+\mu^-$ events.  
The effects on the hadron $\eta ^i_q(x_p)$ are 
small and are included in the error due to charged particle efficiency and 
purity.

\item {\bf Efficiencies of {\boldmath$\rm K^0_S$} and {\boldmath$\Lambda$}:}

The uncertainties of the $\rm K^0_S$ and $\Lambda$ efficiencies
as given for $x_p> 0.2$ in Table~\ref{tab:flowmtx02}, for example,
are taken into account.  Since the relative yields of $\rm K^0_S$ and 
$\rm K^\pm$ are important for the separation of up and down quark 
jets, the uncertainty contributes significantly also to the 
$\eta _q^{\rm K^\pm}$, for example, to $\rm \eta_u^{K^\pm}$ and 
$\rm \eta_s^{K^\pm}$.  The relevant sources of systematic error are 
described in~\cite{bib-OPALK0S} and~\cite{bib-OPALLAMBDA}.  For the
$\rm K^0_S$ the systematic errors for the region $0.7<|\cos\theta|<0.9$ 
were taken to be double those in the barrel region, motivated by the
factor of two worse mass resolution in the endcap.

\item {\bf Charm tag efficiency:}

The relative uncertainty in the $\rm D^{*\pm}$ reconstruction efficiency was 
conservatively estimated to be $\pm 10\%$.  This source of error has a 
negligible effect on the results.

\item {\bf Hemisphere correlations:}

Correlations due to kinematic and geometrical effects are accounted for by the 
$\rho_{ij}$ parameters, which are taken from Monte Carlo simulation. The values 
of $\rho_{ij}$ are most sensitive to changes in the angular acceptance of the 
tagging particles and the thrust angle cut.  Variations in maximum 
$|\cos\theta|$ of the tagging particles between $0.7$-$0.9$, and different 
cuts on the maximum $|\cos\theta_{\rm T}|$ between $0.7$-$0.9$ show that the 
changes of the $\rho_{ij}$ are well simulated. A $\pm 0.01$ absolute systematic
error, representing the maximal disagreement between data and Monte
Carlo, is assigned for the simulation of the $\rho_{ij}$ values.

\item {\bf Other background:}

Contributions to the detector level $\eta ^i_q$ from events which are not 
tagged at the generator level are taken from the JETSET Monte Carlo events.  
Such events are mainly due to tags which have a true $x_p$ lower than the 
minimum $x_p$ cut used but are tagged due to the finite momentum resolution in
the detector, spurious tracks, and combinatoric background in the case of the 
$\rm K^0_S$ and $\Lambda$ samples.  Another source of other background (mainly 
in the proton sample) is due to stable charged hyperons, mostly $\Sigma^-$.

These backgrounds represent either an absolute contribution to 
$\eta ^{i{\rm{,\ exp}}}_q(x_p)$ or a constant background fraction which scales
with the $\eta ^i_q(x_p)$.  The systematic errors on the estimations of these 
backgrounds are taken as the differences in the $\eta ^i_q(x_p)$ if the 
analysis is repeated under the two assumptions, namely treating the other 
background events as a fraction of the detector level 
$\eta ^{i{\rm{,\ exp}}}_q(x_p)$ or as an absolute contribution.  This procedure 
takes into account uncertainties in the JETSET modelling of both the magnitude 
and the $x_p$ dependence of the background sources.

\item {\bf Charm tag background:}

The flavour composition of the $\rm D^{*\pm}$ sample has been discussed
in~\cite{bib-OPALD*}.  The fraction of bottom quark jets in the $\rm D^{*\pm}$ 
sample can be directly determined. The contribution from gluon splitting 
$g\rightarrow {\rm c \bar c}$ is negligible. The flavour composition of the 
combinatorial background is taken from Monte Carlo.  We varied the 
contributions individually by $\pm 50\%$.  The corresponding uncertainties have 
been included in the systematic errors.  

\item {\bf Fixed quantities:}

The quantities which were fixed in the fit, namely
$\eta_{\rm d,u,s}^{\mu^\pm}$, $\eta_{\rm d,u,s}^{\rm D^{*\pm}}$, and 
$\eta_{\rm d,u,s}^{\rm b-vtx}$, are each in turn varied by $\pm 100\%$
and the corresponding shifts in the $\eta ^i_q$ taken as systematic errors.

\item {\bf Hadronisation symmetries:}  

As discussed in Section~\ref{sec:method} the hadronisation symmetries used to 
solve the equation system may be broken by up to 2\% at low $x_p$ and 
10\% for the relation between $\eta_{\rm d}^\Lambda(x_p)$ and 
$\eta_{\rm u}^\Lambda(x_p)$~\cite{bib-letmaett,bib-JETSET}.  The relations 
are corrected for any breaking which is present in the JETSET Monte Carlo.  
Assuming a systematic error equal to the maximal allowed breaking, 
the $\eta ^i_q(x_p)$ change only marginally.

\item {\boldmath $\rm Z^0$ \bf branching ratios:} 

The uncertainties due to the $\rm Z^0$ branching ratios $R_q$ into quarks have 
been estimated by varying each fraction within certain limits.  In the case of 
bottom and charm quarks these are given by the rather precise measurements at 
LEP~\cite{bib-PDG}.  Branching fractions into the individual light quarks are 
less well determined.  A direct measurement has only been performed 
in~\cite{bib-ewOPAL}, which we do not consider because it used a variation 
of the method applied in this paper.  However, there are constraints on 
the electroweak couplings of up and down quarks coming from lepton nucleon 
scattering and final-state photon radiation from quarks~\cite{bib-PDG}
which agree with the Standard Model expectation.

To take into account uncertainties in the light flavour $R_q$, we vary 
$R_{i}/(R_{\rm d}+R_{\rm u}+R_{\rm s})$ where $i=\rm d,u,s$ by $\pm 10\%$ 
from the Standard Model values taking into account their well measured sum 
$1-R_{\rm b}-R_{\rm c}=0.606\pm0.010$~\cite{bib-PDG}.  The $\eta ^i_q(x_p)$ 
values change by a maximum of 0.5\%, and the $\chi^2$ only marginally.  Since 
we assume the Standard Model in this analysis, we do not include this small 
source of error in the overall systematic errors.

\end{itemize}

\subsection{Cross-check on events with a heavy quark tag}\label{ssec:claudio}

We followed the procedure detailed in~\cite{bib-ewOPAL} in order to make a 
cross-check of the principal results.  Compared to the light flavour tags based 
on high $x_p$ stable hadrons, the purity of heavy quark tags is much higher.
The cross-check makes use of the charm and bottom tag efficiencies and purities 
from Monte Carlo, mentioned in Section~\ref{sec:cbtag}.  By counting the 
number of light flavour tags in event hemispheres opposite to a heavy flavour 
tag, one can determine the $\eta_q^{i,\ \rm exp}(x_p)$ directly without using 
a large system of equations.  This method leads to results which are consistent 
with those from the main method used in this paper.  The agreement between data 
and the JETSET model is found to be in general quite good.  The biggest 
discrepancy between the data and the prediction of the JETSET model is for 
$\eta^{\Lambda}_{\rm{c}} (x_p)$, which will be discussed in more detail in 
Section~\ref{sec:compmodel}.


\section{Results and Hadronisation Studies}\label{sec:hadro_study}

The $\eta ^i_q(x_p)$ values with their statistical and systematic uncertainties 
are listed in Table~\ref{tab:results02} for $x_p > 0.2$.  The largest 
$\eta ^i_q$ in light flavours for mesons are shown in 
Figures~\ref{fig:tag-eff-int-mes1} and~\ref{fig:tag-eff-int-mes2} as 
a function of the cut on $x_{p}$.  In addition, for baryons, the 
${\rm \eta ^{p}_{\rm u}}$, ${\rm \eta ^{p}_{\rm d}}$, 
${\rm \eta ^{\Lambda}_{\rm s}}$ and ${\rm \eta ^{\Lambda}_{\rm c}}$ are shown 
in Figures~\ref{fig:baryon_int1} and~\ref{fig:baryon_int2}.

In most cases only weak correlations exist between the $\eta ^i_q$ for 
different tagging particles $i$, though stronger correlations exist between 
different flavour $\eta ^i_q$ with the same tagging hadron.  The statistical
correlation coefficients for the most important $\eta ^i_q$ are given in 
Table~\ref{tab:corr02} for $x_p>0.2$.  These correlations are typical also for 
the other minimum values of $x_p$.

In all cases the expected pattern of the leading particles holds: the up and 
down quarks fragment mostly into pions, whereas the strange quarks fragment 
mostly into $\rm K^\pm$ and $\rm K^0_S$, although the fraction of $\pi^\pm$ is 
sizable also for strange quarks, especially at low $x_p$.  The heavy charm and 
bottom quarks produce mostly high-energy pions and kaons.

\subsection{Comparison to JETSET and HERWIG}\label{sec:compmodel}

\begin{figure}[p]
\begin{center}
\resizebox{\textwidth}{!}
{\includegraphics{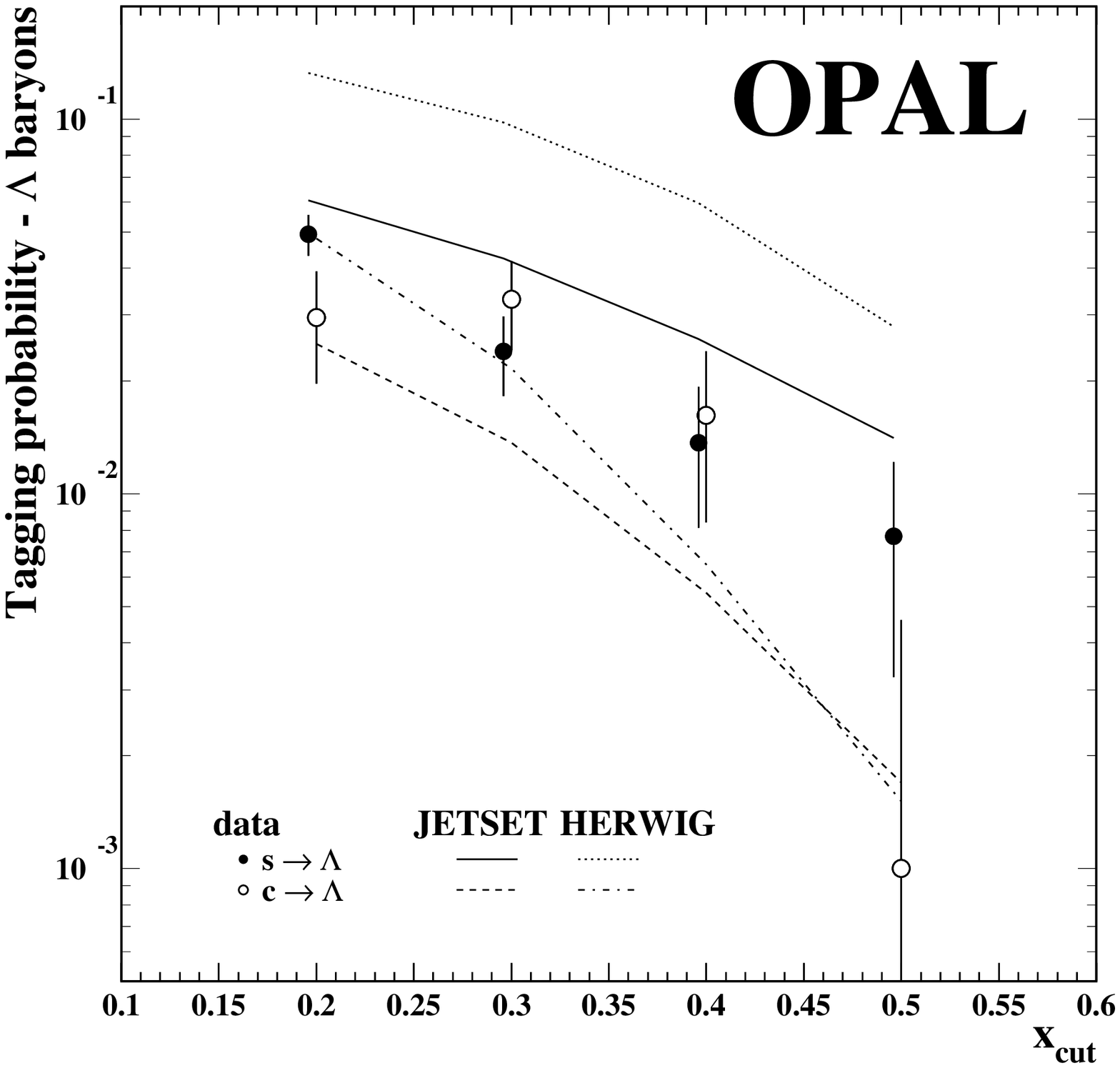}}
\caption{Tagging probabilities as a function of 
the minimum $x_p$ cut for $\Lambda$ baryons.
Data points are correlated for different values of the minimum $x_p$ cut.
The errors shown are statistical plus systematic.
The lines show the JETSET and HERWIG predictions.
\label{fig:baryon_int2} }
\end{center}\end{figure}

We compare the dominant fragmentation functions for the different flavours 
to the expectations of the HERWIG and JETSET models.  The results are shown in 
Figures~\ref{fig:tag-eff-int-mes1} and~\ref{fig:tag-eff-int-mes2} for mesons, 
and in Figures~\ref{fig:baryon_int1} and~\ref{fig:baryon_int2} for baryons.
Note that the data points are shown for different minimum values of the $x_p$ 
and therefore are correlated.  For $x >$ 0.2 the JETSET and HERWIG expectations 
for all determined fragmentation functions are listed together with the data 
in Table~\ref{tab:results02}.

Hadronisation is quite differently modelled in the two QCD generators.
Whereas JETSET uses the Lund string model~\cite{bib-Lundstring},
HERWIG invokes principally the cluster decay mechanism~\cite{bib-Wolfram}.
Both models, JETSET more than HERWIG, contain several parameters which 
cannot be derived from first principles.  For this comparison we use the 
standard OPAL tuning~\cite{bib-toon} which is optimised to describe the 
overall event properties and inclusive particle production. Our measurements 
of the flavour dependence of the fragmentation function allows us to test
the correctness of the model at a new level of detail.

Most of the tagging probabilities are well reproduced by both JETSET and 
HERWIG.  Exceptions are the consistent underestimation in HERWIG of the meson
production in bottom events.  In addition HERWIG seems to underestimate 
$\eta ^{\pi}_{\rm s}$ and both HERWIG and JETSET seem to underestimate 
$\eta ^{\rm K^{\pm}}_{\rm u}$.  The significance of these deviations is, 
however, only at the level of two standard deviations.  The distributions for 
kaons have a similar shape for the various quark species.  The distributions 
for pions are significantly steeper than those for kaons, which can at least 
partly be explained by the larger fraction of pions from resonance decays that 
will be found at lower $x_p$ values.  Particularly $\eta^\pi_{\rm s}$, which 
can only be due to either decays or if both an up and a down quark are produced 
from the hadronisation sea, is steeper than $\eta^{\rm K}_{\rm s}$ and 
$\eta^\pi_{\rm d}$.

Whereas proton production is reasonably described by JETSET, the HERWIG 
prediction deviates significantly from the data.  The flavour integrated 
rate~\cite{bib-protonrates} is overestimated at high $x_p$ in this model.  As 
can be seen from Table~\ref{tab:results02}, especially the fractions of protons 
in up and down quark events are higher.  In fact the excess in the HERWIG
prediction is almost exclusively due to these quarks, which are in HERWIG for 
$x_p>0.2$ about twice as high as in the data.  The data also show that the 
yield of protons from up quarks is higher than that from down quarks.  This 
will be discussed in more detail in Section~\ref{sec:baryon}.  

\begin{figure}[p]
\begin{center}
\resizebox{\textwidth}{!}
{\includegraphics{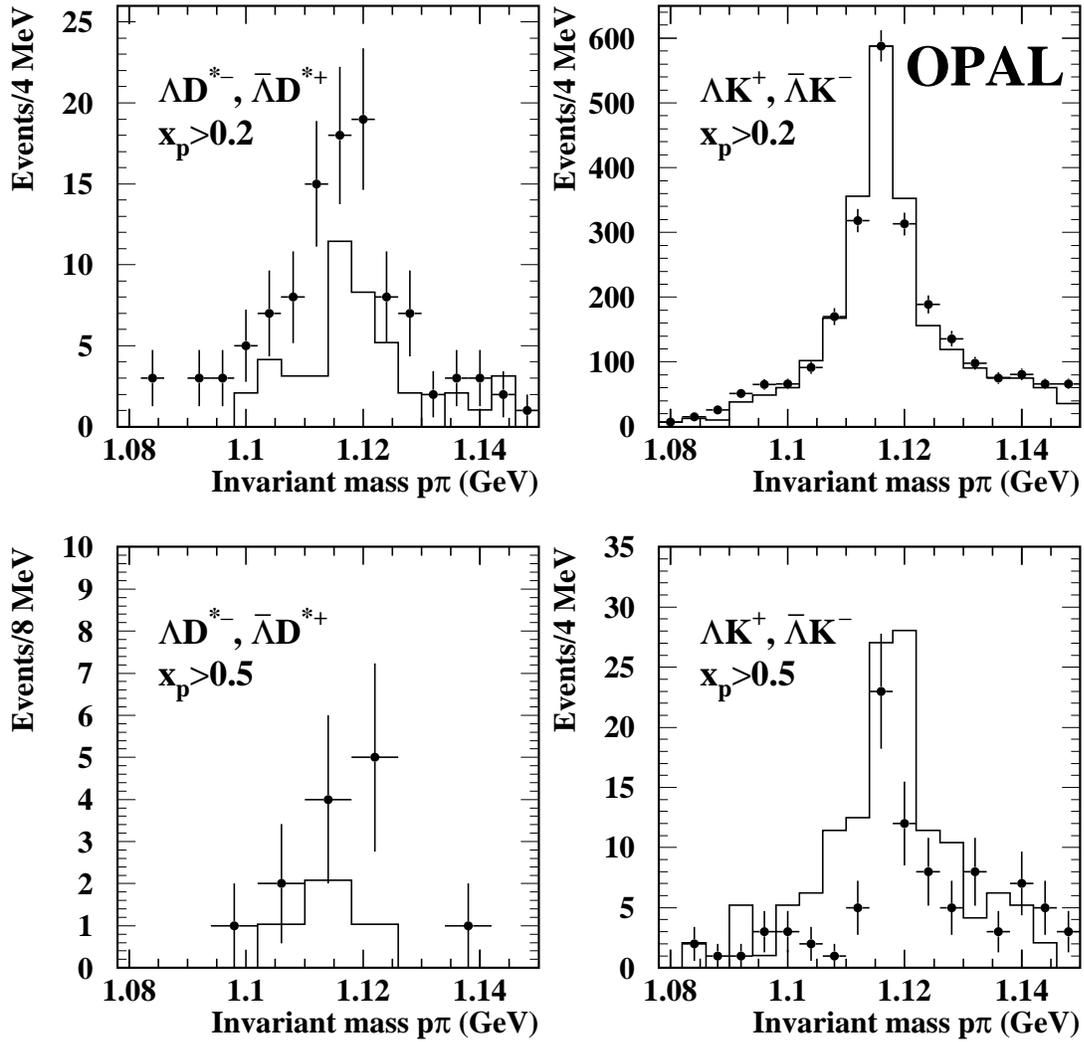}}
\caption{Invariant mass distributions for the $\Lambda$ signals in data 
(points with error bars) and the JETSET Monte Carlo (histogram).  The Monte Carlo
is normalised to the same number of events with a $\rm D^{*-}$ (left) or a 
$\rm K^+$ (right) in the opposite hemisphere, for two different $x_p$ ranges.
\label{fig:c-to-lambda}}
\end{center}
\end{figure}

HERWIG also significantly overestimates $\Lambda $ production for all light 
quark species, as can be seen in Figure~\ref{fig:baryon_int2}.  The 
overestimation for $\eta ^{\Lambda}_{\rm c}(x>0.2)$ is less pronounced; 
however, the shape of the fragmentation function is softer than in the data.  
In the case of $\Lambda$ baryons in strange and charm events, the JETSET 
expectation differs from the data as seen in Figure~\ref{fig:baryon_int2}.  
The $\rm s\rightarrow \Lambda $ yield in the data is only about half of that 
expected although the shape is consistent.  For $\rm c\rightarrow \Lambda$ the 
yield is underestimated by a factor of 2$-$3 and the $x$-dependence tends to 
be steeper.  To study whether the discrepancy in the rate may be due to the 
analysis procedure, we compare $\Lambda$ 
production directly in data and in the JETSET simulation including detector 
effects. To enrich charm and strange events, respectively, we search for 
$\Lambda$ in hemispheres opposite to a tagged $\rm D^{*-}$ or $\rm K^+$, and 
$\overline\Lambda$ production in hemispheres opposite to a tagged $\rm D^{*+}$ 
or $\rm K^-$.  The resulting $\rm p\pi^-$ mass spectra are shown in 
Figure~\ref{fig:c-to-lambda}.  The underestimation of the $\Lambda$ production 
in charm events in the simulation is clearly visible, as is the overestimation 
at high $x_p$ of $\Lambda$ production in strange events.

In addition to studying absolute rates of individual particle species
for a specific flavour, as the next step we compare relative yields for 
the same flavour or the same particle type.  These relations may reveal 
symmetries in the hadronisation mechanism.

\subsection{Strange quark fraction in QCD vacuum}

\begin{figure}[p]
\begin{center}
\resizebox{\textwidth}{!}
{\includegraphics{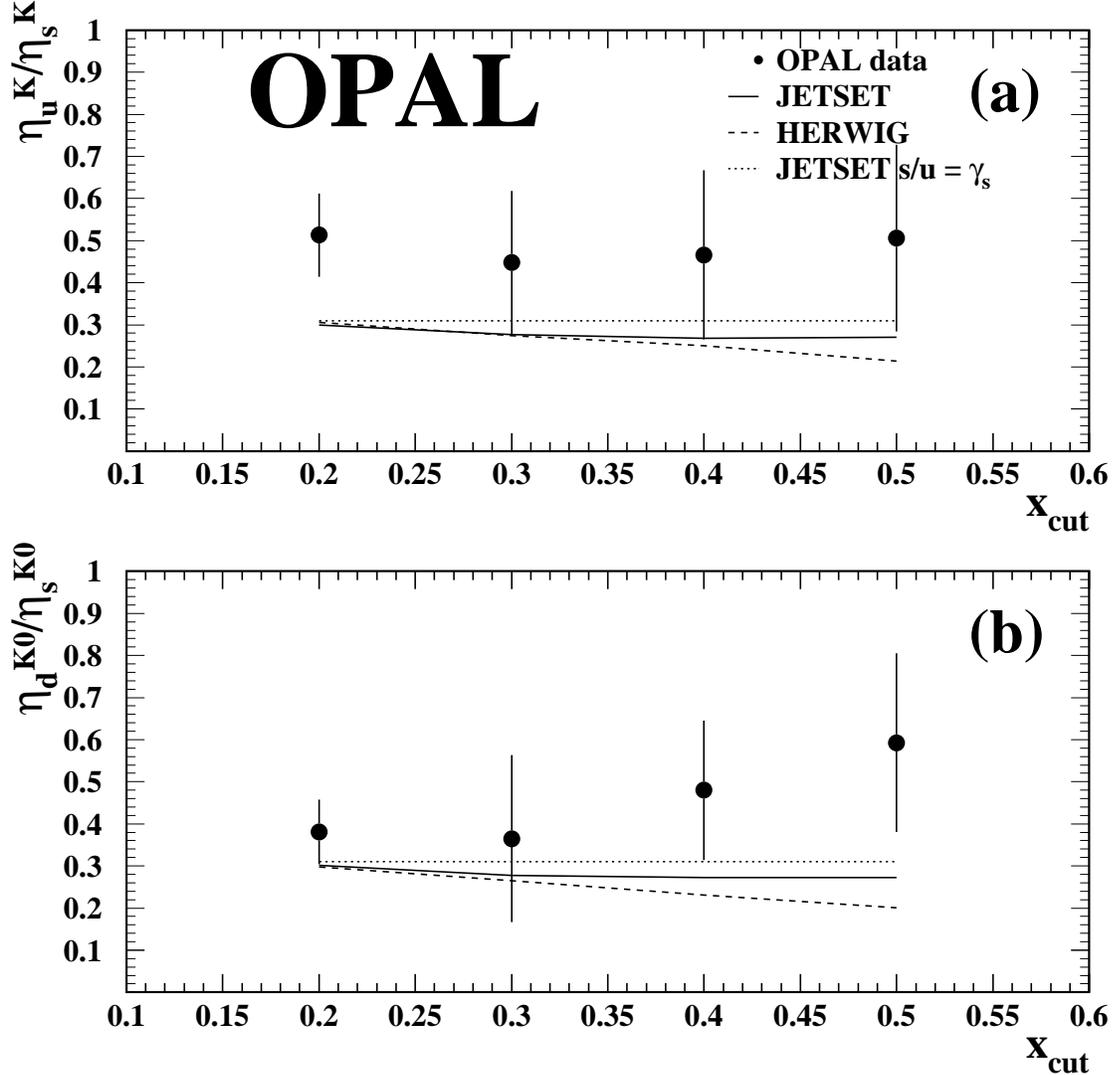}}
\caption{Determination of $\gamma_{\rm s}={\rm s/u}$ using the 
estimator $\eta_{\rm u}^{\rm K^\pm} / \eta_{\rm s}^{\rm K^\pm}$ (a) 
and $\eta_{\rm d}^{\rm K^0_S} / \eta_{\rm s}^{\rm K^0_S}$ in the (b)
in the OPAL data (solid points).  Data points are correlated for different 
values of the minimum $x_p$ cut and the errors shown are statistical plus 
systematic.  The solid lines represent the true
$\eta_{\rm u}^{\rm K^\pm} / \eta_{\rm s}^{\rm K^\pm}$ and
$\eta_{\rm d}^{\rm K^0_S} / \eta_{\rm s}^{\rm K^0_S}$ in 
the JETSET Monte Carlo and the dashed lines the HERWIG predictions. 
The dotted lines represent the input value of 
$\gamma_{\rm s}={\tt PARJ(2)}=0.31$ in JETSET.  
\label{fig:gammaskaon}}
\end{center}
\end{figure}

In a next step we compare the yield of $\rm K^\pm$ in up and strange quark 
events and $\rm K^0_S$ in down and strange events.  Within JETSET the ratio 
of the production yields of the primary hadrons is a direct measure of 
$\gamma_{\rm s} = {\cal P}({\rm s}) / {\cal P}({\rm u,d})$, i.e.\ the relative 
quark production probabilities in the hadronisation sea.  We present
the results in Figure~\ref{fig:gammaskaon} and Table~\ref{tab:gammas}.  The 
full lines show the expected ratio in JETSET for a $\gamma _s$ value indicated 
by the dotted lines.  The difference of up to 10\% between the expected ratio 
and $\gamma_{\rm s}$ is due to decays, particularly of the $L=1$ meson
supermultiplet.  The comparisons show that
$\rm \eta_{\rm u}^{K^{\pm}}/\eta_{\rm s}^{K^{\pm}}$ 
(Figure~\ref{fig:gammaskaon}a) and
$\rm \eta_{\rm d}^{K^0}/\eta_{\rm s}^{K^0}$ 
(Figure~\ref{fig:gammaskaon}b) are good estimators of $\gamma_{\rm s}$. 

No significant dependence on $x_p$ is observed for either the $\rm K^\pm$ or
$\rm K^0_S$ measurements, which is consistent with expectations.  A combined 
$\rm K^\pm$ and $\rm K^0_S$ analysis is made by invoking SU(2) isospin 
symmetry which implies that $ \rm \eta_u^{K^\pm} = \eta_d^{K^0}$.
After taking into account correlations and correcting for isospin-violating 
decays at $x_p>0.2$, we obtain 
$$ \rm \gamma _s  \ = \ 0.422 \pm 0.049 (stat.) \pm 0.059 (syst.) $$
The systematic uncertainty includes an error of 0.042 to take into account 
variations of the correction factors due to the uncertain amounts
of resonance production, found by varying the contributions of the
$L=1$ meson supermultiplet by $\pm 50\%$.  

This value of $\rm \gamma_s$ is consistent with, although somewhat larger 
than, previous measurements~\cite{bib-gammasrev} which are,
however, in most cases rather indirect.  Comparing the data in more 
detail with the JETSET prediction in Table~\ref{tab:results02}, one observes 
that the $\rm \eta _s^{K^\pm}$ and $\rm \eta _s^{K^0_S}$ are in good agreement.
However, in the data more ${\rm K^\pm}$ are found in up quark events
and more ${\rm K^0_S}$ in down quark events than predicted by JETSET.

In the case of HERWIG the ratios of the flavour dependent ${\rm K^0_S}$ and
${\rm K^\pm}$ production have no simple interpretation in terms of a
single parameter. The ratios are fairly similar to those of JETSET and thus 
agree with the data.

\subsection{Baryon hadronisation}\label{sec:baryon}

The mechanism of how three quarks coalesce in the jet development to form a 
baryon is still a puzzle of hadronisation.  Our measurement of the flavour 
dependence of the proton and $\Lambda $ yields provides additional new input.

\begin{figure}[p]
\begin{center}
\resizebox{\textwidth}{!}
{\includegraphics{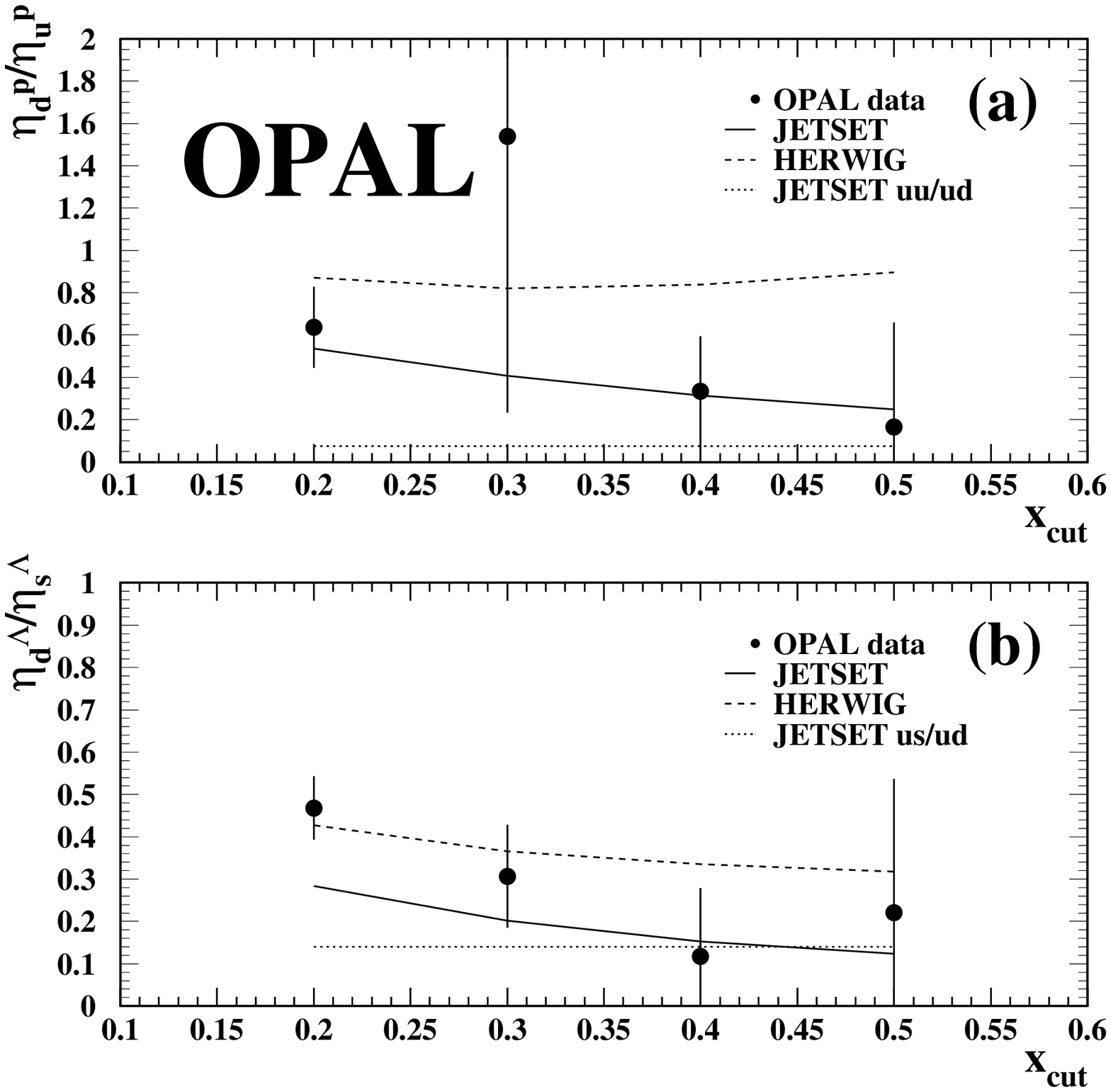}}
\caption{
$\eta_{\rm d}^{\rm p} / \eta_{\rm u}^{\rm p}$ (a) and
$\eta_{\rm d}^{\Lambda} / \eta_{\rm s}^{\Lambda}$ (b)
in the OPAL data (solid points).  
Data points are correlated for different values of the minimum $x_p$ cut
and the errors shown are statistical plus systematic.
The solid lines represent the true
$\eta_{\rm d}^{\rm p} / \eta_{\rm u}^{\rm p}$  and
$\eta_{\rm d}^{\rm \Lambda} / \eta_{\rm s}^{\Lambda}$ in 
the JETSET Monte Carlo and the dashed lines the HERWIG predictions. 
The dotted lines represent the input values of 
$\rm uu/ud = 3 \cdot { \tt PARJ(4) } = 0.075  $ (a) and
$\rm us/ud = \gamma_{\rm s}\cdot  { \tt PARJ(3) } = 0.1395 $ (b) in JETSET.
\label{fig:baryonsupp}}
\end{center}
\end{figure}

The ratio $\rm \eta^p_d/\eta^p_u$ is shown\footnote{The results for 
$x_p>0.30$ are not very precise due to a lack of separation power between 
up and down quarks.} in Figure~\ref{fig:baryonsupp}a and listed in 
Table~\ref{tab:baryonsupp}.  Within the LUND string model ideally the ratio 
$\rm \eta^p_d/\eta^p_u$ at high $x_p$ would be a direct measure of the size 
of the suppression of diquarks~\cite{bib-diquarks} with spin~1 relative to 
spin~0, since Fermi statistics requires a (uu) diquark to have angular momentum 
$L=1$.  However, decays from heavier baryons such as $\Lambda$ or $\Delta$ 
resonances tend to change the ratio.  Although our result agrees with the 
production of diquarks as suggested in~\cite{bib-diquarks} and already 
supported by studies of baryon number compensation in 
jets~\cite{bib-expdiquarks}, the uncertainties are so large that the data are 
also consistent with models that form baryons from quarks that are
statistically produced in rapidity.
HERWIG, which incorporates a democratic production of diquarks, predicts the 
production of protons from u and d jets to be more equal than JETSET.  The 
data tend to be smaller than the HERWIG expectation.

We also observe the suppression of strangeness in baryon production by 
measuring the ratio of $\Lambda$ baryon production in down and strange quark 
events.  Note that in solving the equation system~(\ref{eq:chi2}) we have 
assumed that $\rm \eta_d^\Lambda \sim \eta_u^\Lambda$.  After the production of 
the primary down or strange quark, $\Lambda$ baryons are formed by picking up a 
pair of (us) or (ud) quarks, respectively.  Indeed fewer $\Lambda$ baryons are 
found in down (and hence up) quark jets than in strange jets.  The suppression 
agrees with the JETSET and HERWIG models.  The results are shown in 
Figure~\ref{fig:baryonsupp}b and listed in Table~\ref{tab:baryonsupp}.

\begin{figure}[p]
\begin{center}
\resizebox{\textwidth}{!}
{\includegraphics{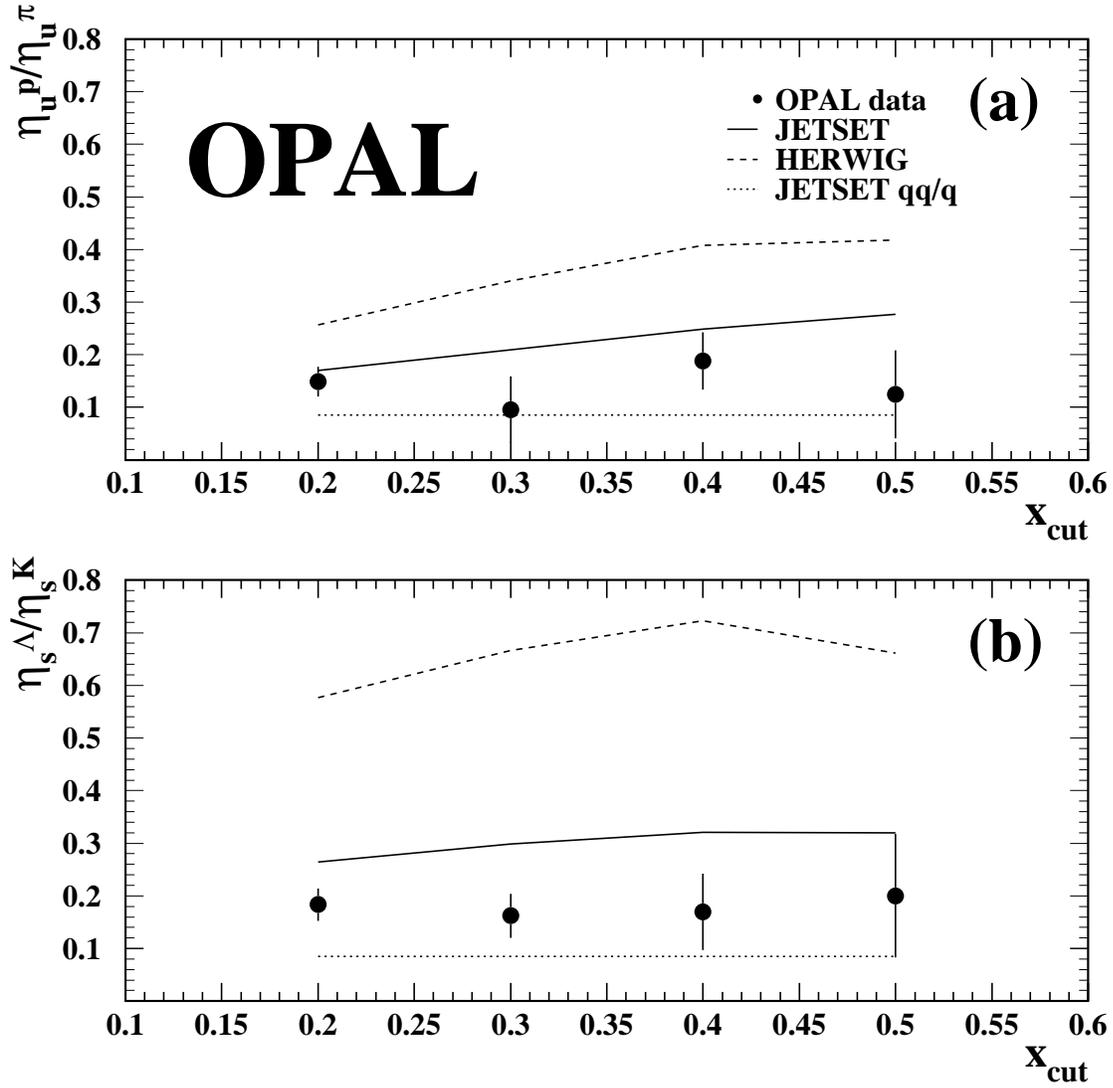}}
\caption{
$\eta_{\rm u}^{\rm p} / \eta_{\rm u}^{\pi}$ (a) and
$\eta_{\rm s}^{\Lambda} / \eta_{\rm s}^{\rm K}$ (b)
in the OPAL data (solid points).  
Data points are correlated for different values of the minimum $x_p$ cut
and the errors shown are statistical plus systematic.
The solid lines represent the true
$\eta_{\rm u}^{\rm p} / \eta_{\rm u}^{\pi^\pm}$  and
$\eta_{\rm s}^{\rm \Lambda} / \eta_{\rm s}^{\rm K^\pm}$ in 
the JETSET Monte Carlo and the dashed lines the HERWIG predictions. 
The dotted lines show the input values for 
$\rm qq/q = { \tt PARJ(1) } = 0.085$ in JETSET.
\label{fig:baryonmeson}}
\end{center}
\end{figure}

Finally we compare the production of baryons and mesons in events of
the same primary quark type.  The ratio of proton to pion production
in up quark events and the ratio of $\Lambda$ to charged kaon
production in strange events are given in Table~\ref{tab:baryonmeson}
and shown in Figures~\ref{fig:baryonmeson}a and~\ref{fig:baryonmeson}b, 
respectively. The JETSET expectations fall above the measured data points.  
Although these measured ratios are poor estimators of the level of diquark 
suppression (indicated by the dotted line) within the JETSET model due to large 
contributions from decays, the suppression of baryons relative to mesons is 
clearly observed.  For both ratios the HERWIG expectation is significantly 
above the measurement as already mentioned in Section~\ref{sec:compmodel}.

In addition one can form the double ratio, 
$(\eta^{\Lambda }_{\rm s}/\eta^{\rm K}_{\rm s}) /
 (\eta^{\rm p}_{\rm u}   /\eta^{\pi}_{\rm u}) $
which within the JETSET model should measure the same quantity
${\cal P}(\rm ud)/{\cal P}(\rm u)$, modified only by decays, 
where ${\cal P}(x)$ indicates the probability to pick out either a quark or 
diquark $x$ from the QCD vacuum.  The data yield for $x_p>$ 0.2 
$$
\frac{\eta ^{\Lambda }_{\rm s}/\eta ^{\rm K}_{\rm s}} 
     {\eta ^{\rm p}_{\rm u}/\eta ^{\pi}_{\rm u}}  \ = \
      1.23\pm  0.31, 
$$
consistent with the JETSET expectation of 1.55, but significantly lower than
the HERWIG prediction of 2.24.  This indicates that the inclusive 
baryon production is badly modelled and also the relations between 
different meson or baryon species are unsatisfactorily simulated in HERWIG.


\section{Conclusions}

In this paper we have reported on a determination of the probabilities
$\eta ^i_q(x_p)$ of leading particles to originate from individual quark
flavours in $\rm Z^0$ decays.  We studied the production of leading
$\pi^\pm$, $\rm K^\pm$, $\rm K^0_S$, proton, and $\Lambda$
for $x_p>0.2$ up to 0.5.  The measurement has only a
minimal reliance on hadronisation models.  In general we observe the
expected behaviour that the flavour of the primary quark is reflected
in the leading particle, i.e.\ up and down quarks lead mainly to highly
energetic pions, while strange quarks lead mainly to kaons.

These measurements allow several aspects of hadronisation to be studied
rather directly, in contrast to many previous analyses which rely
strongly on a model unfolding of different contributions.  In particular
we determine from the relative production of leading charged kaons in
up and strange quark jets and leading $\rm K^0_S$ in down and strange quark
jets the suppression of strange quarks in the QCD vacuum:
$$ \rm \gamma _s  \ = \ 0.422 \pm 0.049 (stat.) \pm 0.059 (syst.) $$
We also find that leading protons are more frequent in up than
in down quark jets.  We also observe the suppression of strange diquarks in 
$\rm (d,u)\to\Lambda$ events and baryons relative to mesons in events of the 
same quark flavour.

For most quark flavours and particle types the JETSET model reproduces the
measurements well.  A possible exception is the production of $\Lambda$
baryons in charm quark events which appears to have a higher yield and 
a harder fragmentation function than expected.  HERWIG provides in general a 
good description of mesons in light quark events but has deficiencies in 
baryon production, in particular the relative yields of different baryon types 
and the ratios of baryons and mesons in the same flavour jet.

In addition to these hadronisation studies, our measurements of
the $\eta ^i_q(x_p)$ may also be interesting for future experiments.
In providing tagging probabilities for light flavours with hardly
any reliance on hadronisation models, the $\eta ^i_q(x_p)$ can be applied
at other centre-of-mass energies or in the study of heavy particle decays.
This allows a determination of the light flavour production yields and
properties in a model-independent way also for environments other than the 
$\rm Z^0$.


\clearpage
\newpage
\appendix
\par
Acknowledgements:
\par
We particularly wish to thank the SL Division for the efficient operation
of the LEP accelerator at all energies
 and for their continuing close cooperation with
our experimental group.  We thank our colleagues from CEA, DAPNIA/SPP,
CE-Saclay for their efforts over the years on the time-of-flight and trigger
systems which we continue to use.  In addition to the support staff at our own
institutions we are pleased to acknowledge the  \\
Department of Energy, USA, \\
National Science Foundation, USA, \\
Particle Physics and Astronomy Research Council, UK, \\
Natural Sciences and Engineering Research Council, Canada, \\
Israel Science Foundation, administered by the Israel
Academy of Science and Humanities, \\
Minerva Gesellschaft, \\
Benoziyo Center for High Energy Physics,\\
Japanese Ministry of Education, Science and Culture (the
Monbusho) and a grant under the Monbusho International
Science Research Program,\\
Japanese Society for the Promotion of Science (JSPS),\\
German Israeli Bi-national Science Foundation (GIF), \\
Bundesministerium f\"ur Bildung, Wissenschaft,
Forschung und Technologie, Germany, \\
National Research Council of Canada, \\
Research Corporation, USA,\\
Hungarian Foundation for Scientific Research, OTKA T-029328, 
T023793 and OTKA F-023259.\\


\newpage

\begin{table}[p]
\begin{center}
\renewcommand{\arraystretch}{1.2}
\begin{tabular}{||c||c|c|c|c|c|c|c|c|c||} \hline\hline
Assigned & \multicolumn{9}{c||}{True} \\
& $\pi^\pm$ & $\rm K^\pm$ & $\rm p(\overline{p})$ & $\rm e^\pm$ & $\mu^\pm$
& $\rm K^0_S$ & $\Lambda(\overline\Lambda)$ & $\rm D^{*\pm}$ & other \\
\hline \hline
$\pi^\pm$             &
    0.790 &     0.062 &     0.003 &     0.013 &     0.007 &     0.038 &     0.005 &     0.062 &     0.019 \\ \hline
$\rm K^\pm$           &
    0.146 &     0.568 &     0.148 &     0.002 &     0.002 &     0.017 &     0.026 &     0.071 &     0.020 \\ \hline
$\rm p(\overline{p})$ &
    0.040 &     0.246 &     0.551 &     0.002 &     0.001 &     0.014 &     0.081 &     0.036 &     0.031 \\ \hline
$\rm e^\pm$           &
    0.186 &     0.023 &     0.002 &     0.620 &     0.000 &     0.024 &     0.006 &     0.128 &     0.011 \\ \hline
$\mu^\pm$             &
    0.100 &     0.061 &     0.002 &     0.002 &     0.643 &     0.017 &     0.007 &     0.153 &     0.015 \\ \hline
$\rm K^0_S$           &
    0.081 &     0.030 &     0.007 &     0.004 &     0.001 &     0.691 &     0.026 &     0.101 &     0.060 \\ \hline
$\Lambda(\overline\Lambda)$ &
    0.047 &     0.024 &     0.024 &     0.003 &     0.001 &     0.128 &     0.696 &     0.032 &     0.045 \\ \hline
$\rm D^{*\pm}$             &
    0.143 &     0.074 &     0.019 &     0.007 &     0.006 &     0.016 &     0.012 &     0.699 &     0.024 \\ \hline\hline
efficiency            &
0.487 & 0.441 & 0.292 & 0.228 & 0.702 & 0.155 & 0.135 & 0.033 & \\
\hline
\hline
\end{tabular}
\caption{Fractional compositions of the identified samples (rows) in terms
of the true tagging particle, for $x_p>0.2$.  The dominant component
of other tagged events (last column) is tagging particles which pass the 
minimum $x_p$ requirement in the detector but whose true momenta are lower.  
The sum of the elements in each row is one.  The last row gives the average 
efficiency to correctly tag a hemisphere, as taken from Monte Carlo simulation.
Errors are discussed in the text.\label{tab:flowmtx02}}
\renewcommand{\arraystretch}{1.0}
\end{center}
\end{table}


\begin{table}[p]
\begin{center}
\hspace*{-15pt}
\begin{tabular}{||c||c|c|c|c|c|c|c|c|c|c||}
\hline\hline
Particle & Tagged &     \multicolumn{9}{c||}{Double-tagged events} \\ 
type   & hemispheres & $\pi^\pm$               &$\rm K^\pm$             &$\rm p(\bar{p})$        &$\rm e^\pm$
&$\mu^\pm$               &$\rm K^0_S$             &$\Lambda(\bar{\Lambda})$&$\rm D^{*\pm}$               &b-vtx
\\ \hline\hline
$\pi^\pm$     &   855043 &     71601 &     77576 &     16287 &      7127 &      7689 &     10425 &      4717 & 2483 &     25074\\
              &   850442 &     72515 &     71411 &     16258 &      7146 &      6637 &      9299 &      4404 & 2238 &     23221\\ \hline
$\rm K^\pm$   &   506538 &  &     25717 &     10120 &      4135 &      4475 &      7248 &      3193 &      1578 & 14376\\
              &   474123 &  &     22742 &      9182 &      3988 &      4081 &      6497 &      3119 &      1526 & 13344\\ \hline
$\rm p(\bar{p})$        &   101415 &  &  &       963 &       789 &       887 &      1375 &       591 &       314 &      2744\\
                        &   100046 &  &  &      1019 &       839 &       815 &      1219 &       583 &       290 &      2784\\ \hline
$\rm e^\pm$             &    54370 &  &  &  &       501 &      1294 &       594 &       235 &       225 &      6219\\
                        &    56235 &  &  &  &       479 &      1186 &       542 &       253 &       219 &      6221\\ \hline
$\mu^\pm$               &    65029 &  &  &  &  &       905 &       674 &       278 &       293 &      8898\\
                        &    60767 &  &  &  &  &       838 &       541 &       274 &       262 &      8254\\ \hline
$\rm K^0_S$             &    71218 &  &  &  &  &  &       523 &       454 &       239 &      2074\\
                        &    64290 &  &  &  &  &  &       423 &       440 &       194 &      1826\\ \hline
$\Lambda(\bar{\Lambda})$&    31721 &  &  &  &  &  &  &       107 &       111 &      1026\\
                        &    30676 &  &  &  &  &  &  &        95 &        82 &       973\\ \hline
$\rm D^{*\pm}$               &    17432 &  &  &  &  &  &  &  &        57 &       805\\
                        &    16692 &  &  &  &  &  &  &  &        76 &       791\\ \hline
b-vtx                   &   245451 &  &  &  &  &  &  &  &  &     22472\\
                        &   246766 &  &  &  &  &  &  &  &  &     23047\\ \hline
\hline
\end{tabular}
\caption{Number of tagged event hemispheres and double-tagged events for
$x_p>0.2$. The upper numbers are for data and the lower for Monte
Carlo,
normalised to the same numbers of ${\rm Z}\to q\bar{q}$ events.
\label{tab:evcount_x02}
}
\end{center}
\end{table}


\begin{table}[p]
\vspace*{-17mm}
\begin{center}
\begin{tabular}{||l|c||c|c||}
\hline \hline
 & OPAL data & JETSET & HERWIG \\
\hline \hline
$\eta_{\rm d}^{ \pi^\pm }$
               & 0.3866 $\pm$  0.0027 $\pm$  0.0257 &  0.3926  & 0.3558 \\ \hline
$\eta_{\rm u}^{ \pi^\pm }$
               & 0.3831 $\pm$  0.0026 $\pm$  0.0256 &  0.3891  & 0.3558 \\ \hline
$\eta_{\rm s}^{ \pi^\pm }$
               & 0.1701 $\pm$  0.0062 $\pm$  0.0139 &  0.1884  & 0.1367 \\ \hline
$\eta_{\rm c}^{ \pi^\pm }$
               & 0.1728 $\pm$  0.0097 $\pm$  0.0186 &  0.1508  & 0.1480 \\ \hline
$\eta_{\rm b}^{ \pi^\pm }$
               & 0.1350 $\pm$  0.0020 $\pm$  0.0093 &  0.1226  & 0.1129 \\ \hline
\hline
$\eta_{\rm d}^{ \rm K^\pm }$
               & 0.0617 $\pm$  0.0102 $\pm$  0.0080 &  0.0517  & 0.0451 \\ \hline
$\eta_{\rm u}^{ \rm K^\pm }$
               & 0.1227 $\pm$  0.0136 $\pm$  0.0244 &  0.0687  & 0.0703 \\ \hline
$\eta_{\rm s}^{ \rm K^\pm }$
               & 0.2390 $\pm$  0.0056 $\pm$  0.0187 &  0.2294  & 0.2300 \\ \hline
$\eta_{\rm c}^{ \rm K^\pm }$
               & 0.0952 $\pm$  0.0100 $\pm$  0.0208 &  0.1123  & 0.1107 \\ \hline
$\eta_{\rm b}^{ \rm K^\pm }$
               & 0.0623 $\pm$  0.0017 $\pm$  0.0066 &  0.0530  & 0.0464 \\ \hline
\hline
$\eta_{\rm d}^{ \rm p}$
               & 0.0362 $\pm$  0.0063 $\pm$  0.0075 &  0.0356  & 0.0795 \\ \hline
$\eta_{\rm u}^{ \rm p}$
               & 0.0569 $\pm$  0.0086 $\pm$  0.0109 &  0.0666  & 0.0913 \\ \hline
$\eta_{\rm s}^{ \rm p}$
               & 0.0328 $\pm$  0.0042 $\pm$  0.0084 &  0.0232  & 0.0319 \\ \hline
$\eta_{\rm c}^{ \rm p}$
               & 0.0246 $\pm$  0.0058 $\pm$  0.0071 &  0.0266  & 0.0334 \\ \hline
$\eta_{\rm b}^{ \rm p}$
               & 0.0232 $\pm$  0.0010 $\pm$  0.0045 &  0.0253  & 0.0199 \\ \hline
\hline
$\eta_{\rm d}^{ \rm K^0_S}$
               & 0.0461 $\pm$  0.0087 $\pm$  0.0061 &  0.0350  & 0.0345 \\ \hline
$\eta_{\rm u}^{ \rm K^0_S}$
               & 0.0228 $\pm$  0.0107 $\pm$  0.0100 &  0.0251  & 0.0229 \\ \hline
$\eta_{\rm s}^{ \rm K^0_S}$
               & 0.1210 $\pm$  0.0028 $\pm$  0.0096 &  0.1161  & 0.1160 \\ \hline
$\eta_{\rm c}^{ \rm K^0_S}$
               & 0.0586 $\pm$  0.0072 $\pm$  0.0132 &  0.0457  & 0.0505 \\ \hline
$\eta_{\rm b}^{ \rm K^0_S}$
               & 0.0273 $\pm$  0.0012 $\pm$  0.0022 &  0.0226  & 0.0195 \\ \hline
\hline
$\eta_{\rm d}^{ \Lambda}$
               & 0.0231 $\pm$  0.0025 $\pm$  0.0020 &  0.0172  & 0.0566 \\ \hline
$\eta_{\rm u}^{ \Lambda}$
               & 0.0211 $\pm$  0.0023 $\pm$  0.0020 &  0.0158  & 0.0542 \\ \hline
$\eta_{\rm s}^{ \Lambda}$
               & 0.0493 $\pm$  0.0046 $\pm$  0.0041 &  0.0607  & 0.1325 \\ \hline
$\eta_{\rm c}^{ \Lambda}$
               & 0.0295 $\pm$  0.0075 $\pm$  0.0063 &  0.0251  & 0.0480 \\ \hline
$\eta_{\rm b}^{ \Lambda}$
               & 0.0180 $\pm$  0.0009 $\pm$  0.0014 &  0.0182  & 0.0170 \\ \hline
\hline
\end{tabular}
\caption{Results for $x_p>0.2$ after corrections for detector efficiency and 
particle misassignment.  The first error shown is statistical and the second 
systematic.  The two rightmost columns show the JETSET and HERWIG expectations 
with the OPAL tunings.\label{tab:results02}}
\end{center}
\end{table}

\begin{table}[p]
\vspace*{-8mm}
\begin{center}
\hspace*{-27pt}
\begin{tabular}{||c||c|c|c|c|c|c|c|c|c|c||}
\hline
\hline
& $\eta_{\rm d}^{ \pi^\pm }$ & $\eta_{\rm s}^{ \pi^\pm }$ &
$\eta_{\rm d}^{ \rm K^\pm }$ & $\eta_{\rm u}^{ \rm K^\pm }$ &
$\eta_{\rm s}^{ \rm K^\pm }$ &
$\eta_{\rm d}^{ \rm p }$ & $\eta_{\rm u}^{ \rm p }$ & $\eta_{\rm d}^{ \rm K^0_S }$ &
$\eta_{\rm s}^{ \Lambda}$    & $\eta_{\rm c}^{ \Lambda}$ \\ \hline \hline
$\eta_{\rm d}^{ \pi^\pm }$   & 1.000 & $-$0.114 & $-$0.117 & $-$0.050 & $-$0.019 & 0.023 & 0.006 & $-$0.058 & $-$0.015 & $-$0.008 \\ \hline 
$\eta_{\rm s}^{ \pi^\pm }$   & & 1.000 & $-$0.013 & $-$0.118 & $-$0.334 & 0.022 & $-$0.088 & 0.081 & $-$0.029 & 0.091 \\ \hline 
$\eta_{\rm d}^{ \rm K^\pm }$ & & & 1.000 & $-$0.813 & 0.250 & $-$0.648 & 0.503 & $-$0.077 & $-$0.081 & 0.067 \\ \hline 
$\eta_{\rm u}^{ \rm K^\pm }$ & & & & 1.000 & $-$0.354 & 0.541 & $-$0.603 & $-$0.009 & 0.127 & $-$0.099 \\ \hline 
$\eta_{\rm s}^{ \rm K^\pm }$ & & & & & 1.000 & $-$0.091 & 0.218 & 0.168 & $-$0.206 & 0.117 \\ \hline 
$\eta_{\rm d}^{ \rm p}$      & & & & & & 1.000 & $-$0.825 & 0.078 & $-$0.015 & 0.086 \\ \hline 
$\eta_{\rm u}^{ \rm p}$      & & & & & & & 1.000 & $-$0.086 & 0.047 & $-$0.013 \\ \hline 
$\eta_{\rm d}^{ \rm K^0_S}$  & & & & & & & & 1.000 & $-$0.018 & 0.082 \\ \hline 
$\eta_{\rm s}^{ \Lambda}$    & & & & & & & & & 1.000 & $-$0.653 \\ \hline 
$\eta_{\rm c}^{ \Lambda}$    & & & & & & & & & & 1.000 \\ \hline 
\hline
\end{tabular}
\caption{Statistical correlations between selected
parameters for $x_p>0.2$.\label{tab:corr02}}
\end{center}
\end{table}


\clearpage
\newpage
\begin{table}[p]
\begin{center}
\hspace*{-13pt}
\begin{tabular}{||l||c@{$\pm$}c|c@{$\pm$}c|c@{$\pm$}c|c@{$\pm$}c|c@{$\pm$}c||} 
\hline\hline
 & \multicolumn{2}{c|}{$x_p>0.2$}
 & \multicolumn{2}{c|}{$x_p>0.3$}
 & \multicolumn{2}{c|}{$x_p>0.4$}
 & \multicolumn{2}{c|}{$x_p>0.5$}
 & \multicolumn{2}{c|}{$x_p>0.6$} \\ \hline\hline
$\eta_{\rm d}^{\pi^\pm}$   &0.3866&0.0258 &0.1924 &0.0130 &0.0888 &0.0062 &0.0400 &0.0031 &0.0183 &0.0030 \\  \hline
$\eta_{\rm u}^{\pi^\pm}$   &0.3831&0.0257 &0.1915 &0.0131 &0.0889 &0.0063 &0.0400 &0.0032 &0.0183 &0.0031 \\  \hline
$\eta_{\rm s}^{\pi^\pm}$   &0.1701&0.0152 &0.0745 &0.0087 &0.0280 &0.0056 &0.0089 &0.0029 &0.0018 &0.0033 \\  \hline
$\eta_{\rm c}^{\pi^\pm}$   &0.1728&0.0210 &0.0652 &0.0113 &0.0228 &0.0068 &0.0079 &0.0038 &0.0013 &0.0056 \\  \hline
$\eta_{\rm b}^{\pi^\pm}$   &0.1350&0.0095 &0.0450 &0.0033 &0.0156 &0.0014 &0.0051 &0.0006 &0.0014 &0.0005 \\  \hline
\hline
$\eta_{\rm d}^{\rm K^\pm}$ &0.0617&0.0129 &0.0257 &0.0152 &0.0071 &0.0086 &0.0011 &0.0047 &0.0007 &0.0179 \\  \hline
$\eta_{\rm u}^{\rm K^\pm}$ &0.1227&0.0280 &0.0664 &0.0265 &0.0376 &0.0157 &0.0195 &0.0080 &0.0048 &0.0204 \\  \hline
$\eta_{\rm s}^{\rm K^\pm}$ &0.2390&0.0195 &0.1480 &0.0116 &0.0807 &0.0079 &0.0385 &0.0045 &0.0209 &0.0041 \\  \hline
$\eta_{\rm c}^{\rm K^\pm}$ &0.0952&0.0231 &0.0405 &0.0130 &0.0150 &0.0072 &0.0070 &0.0039 &0.0010 &0.0045 \\  \hline
$\eta_{\rm b}^{\rm K^\pm}$ &0.0623&0.0069 &0.0190 &0.0024 &0.0045 &0.0009 &0.0007 &0.0004 &0.0000 &0.0008 \\  \hline
\hline
$\eta_{\rm d}^{\rm p}$     &0.0362&0.0098 &0.0281 &0.0100 &0.0056 &0.0036 &0.0008 &0.0023 &0.0000 &0.0085 \\  \hline
$\eta_{\rm u}^{\rm p}$     &0.0569&0.0139 &0.0183 &0.0120 &0.0167 &0.0058 &0.0050 &0.0036 &0.0010 &0.0055 \\ \hline
$\eta_{\rm s}^{\rm p}$     &0.0328&0.0094 &0.0171 &0.0059 &0.0051 &0.0032 &0.0044 &0.0021 &0.0014 &0.0022 \\  \hline
$\eta_{\rm c}^{\rm p}$     &0.0246&0.0092 &0.0076 &0.0047 &0.0047 &0.0032 &0.0006 &0.0014 &0.0008 &0.0033 \\  \hline
$\eta_{\rm b}^{\rm p}$     &0.0232&0.0046 &0.0067 &0.0015 &0.0016 &0.0005 &0.0004 &0.0002 &0.0002 &0.0001 \\  \hline
\hline
$\eta_{\rm d}^{\rm K^0_S}$ &0.0461&0.0106 &0.0271 &0.0144 &0.0193 &0.0065 &0.0114 &0.0034 &0.0041 &0.0162 \\  \hline
$\eta_{\rm u}^{\rm K^0_S}$ &0.0228&0.0146 &0.0126 &0.0177 &0.0027 &0.0081 &0.0000 &0.0034 &0.0000 &0.0206 \\  \hline
$\eta_{\rm s}^{\rm K^0_S}$ &0.1210&0.0100 &0.0743 &0.0059 &0.0402 &0.0040 &0.0192 &0.0022 &0.0103 &0.0020 \\  \hline
$\eta_{\rm c}^{\rm K^0_S}$ &0.0586&0.0151 &0.0289 &0.0084 &0.0099 &0.0055 &0.0043 &0.0041 &0.0009 &0.0040 \\  \hline
$\eta_{\rm b}^{\rm K^0_S}$ &0.0271&0.0025 &0.0088 &0.0013 &0.0020 &0.0005 &0.0002 &0.0003 &0.0000 &0.0001 \\  \hline
\hline
$\eta_{\rm d}^{\Lambda}$   &0.0231&0.0032 &0.0074 &0.0025 &0.0016 &0.0022 &0.0017 &0.0017 &0.0000 &0.0003 \\  \hline
$\eta_{\rm u}^{\Lambda}$   &0.0211&0.0031 &0.0071 &0.0024 &0.0016 &0.0022 &0.0017 &0.0017 &0.0000 &0.0003 \\  \hline
$\eta_{\rm s}^{\Lambda}$   &0.0493&0.0062 &0.0240 &0.0058 &0.0137 &0.0056 &0.0077 &0.0044 &0.0013 &0.0043 \\  \hline
$\eta_{\rm c}^{\Lambda}$   &0.0295&0.0098 &0.0330 &0.0088 &0.0162 &0.0078 &0.0010 &0.0036 &0.0022 &0.0093 \\  \hline
$\eta_{\rm b}^{\Lambda}$   &0.0180&0.0016 &0.0045 &0.0009 &0.0004 &0.0009 &0.0000 &0.0001 &0.0000 &0.0001 \\  \hline
\hline
$\chi^2$ 
& \multicolumn{2}{c|}{40.8}
& \multicolumn{2}{c|}{40.6}
& \multicolumn{2}{c|}{53.4}
& \multicolumn{2}{c|}{39.4}
& \multicolumn{2}{c|}{39.0} \\ \hline
\hline
\end{tabular}
\caption{Results for various values of the minimum $x_p$
requirement with statistical and systematic errors combined.  In the last 
row, the $\chi^2$ of the solution is given.  The number of parameters in 
the fit is 45, of which 32 are free, nine are fixed to their Monte Carlo 
values and four by hadronisation symmetries.
\label{tab:xintegral}}
\end{center}
\end{table}


\begin{table}[p]
\begin{center}
\renewcommand{\arraystretch}{1.2}
\hspace*{-30pt}
\begin{tabular}{||l||c|c|c|c|c|c|c||c||} \hline\hline
Source of Error
& $\rm \eta_{\rm d}^{\pi^{\pm}}$
& $\rm \eta_{\rm s}^{\pi^{\pm}}$
& $\rm \eta_{\rm u}^{K^{\pm}}$
& $\rm \eta_{\rm s}^{K^{\pm}}$
& $\rm \eta_{\rm u}^{p}$
& $\rm \eta_{\rm s}^{\Lambda}$
& $\rm \eta_{\rm c}^{\Lambda}$
& $\gamma_{\rm s}({\rm K^\pm})$
\\ \hline \hline
Charged purity and eff.        &0.0254 &0.0114 & 0.0226 &0.0176 &0.0106 &0.0010 &0.0004 &0.0615 \\ \hline
$\rm K^0_S$ purity and eff.    &0.0002 &0.0006 & 0.0048 &0.0043 &0.0008 &0.0006 &0.0004 &0.0298 \\ \hline
$\Lambda$ purity and eff.      &0.0001 &0.0001 & 0.0005 &0.0005 &0.0007 &0.0029 &0.0017 &0.0011 \\ \hline
Charm tag purity and eff.      &       &       & 0.0001 &       &       &       &0.0001 &0.0003 \\ \hline
Hemisphere correlations        &0.0012 &0.0049 & 0.0050 &0.0033 &0.0024 &0.0024 &0.0057 &0.0293 \\ \hline
Other background               &0.0002 &0.0003 & 0.0008 &0.0002 &0.0002 &0.0003 &0.0001 &0.0032 \\ \hline
Charm tag background           &0.0017 &0.0061 & 0.0033 &0.0028 &0.0006 &0.0011 &0.0019 &0.0121 \\ \hline
Fixed quantities $\to 0$       &0.0002 &0.0008 & 0.0006 &0.0003 &0.0001 &0.0001 &0.0003 &0.0026 \\ \hline
Hadronisation symmetries       &0.0033 &0.0002 & 0.0039 &0.0016 &0.0005 &0.0004 &0.0005 &0.0188 \\ \hline
$\delta R_{\rm c}$             &0.0009 &0.0004 & 0.0006 &0.0008 &0.0001 &0.0001 &0.0002 &0.0043 \\ \hline
$\delta R_{\rm b}$             &0.0001 &0.0001 & 0.0001 &0.0001 &       &       &0.0001 &0.0005 \\ \hline
\hline
Total Systematic Error         &0.0257 &0.0139 & 0.0244 &0.0187 &0.0109 &0.0041 &0.0063 &0.0779 \\ \hline
\hline
\end{tabular}
\caption{Systematic errors on the measurements of the $\eta^i_q$, corrected 
for detector efficiency and particle misassignment, for $x_p>0.2$.  Also shown 
in the last column are the systematic error contributions for 
$\rm \gamma _s(K^\pm)$.  Absence of a number means that the error was less than 
$5\times10^{-5}$.\label{tab:systerr02}}
\renewcommand{\arraystretch}{1.0}
\end{center}
\end{table}

\clearpage
\newpage

\begin{table}[p]
\begin{center}
\begin{tabular}{||l||c|c||}
\hline \hline
$x_{\rm cut}$ &
$\gamma_{\rm s}(\rm K^\pm)=\eta^{\rm K^\pm}_{\rm u}/\eta^{\rm K^\pm}_{\rm s}$ &
$\gamma_{\rm s}(\rm K^0_S)=\eta^{\rm K^0_S}_{\rm d}/\eta^{\rm K^0_S}_{\rm s}$ \\ 
\hline\hline
$x_p>0.2$ &0.513$\pm$0.060$\pm$0.078 &0.381$\pm$0.069$\pm$0.034  \\ \hline
$x_p>0.3$ &0.448$\pm$0.117$\pm$0.123 &0.365$\pm$0.127$\pm$0.152  \\ \hline
$x_p>0.4$ &0.466$\pm$0.140$\pm$0.144 &0.480$\pm$0.141$\pm$0.087  \\ \hline
$x_p>0.5$ &0.506$\pm$0.172$\pm$0.140 &0.593$\pm$0.140$\pm$0.160  \\ \hline
\hline
$x_p>0.2$ & \multicolumn{2}{c|}{$0.422 \pm 0.049 \pm 0.059$} \\ \hline
\hline
\end{tabular}
\caption{Results for different values of the minimum $x_p$ cut
for strange quark suppression as estimated by
$\rm \gamma_s(K^\pm)=\eta^{K^\pm}_{u}/\eta^{K^\pm}_{\rm s}$ and
$\rm \gamma_s(K^0_S)=\eta^{K^0_S}_{d}/\eta^{K^0_S}_{\rm s}$
with statistical and systematic errors calculated taking into account
correlations between the numerators and denominators of the ratios. 
In the last row a combined charged and neutral kaon result is given,
corrected for decays.
\label{tab:gammas}}
\end{center}
\end{table}

\begin{table}[p]
\begin{center}
\begin{tabular}{||l||c|c||}
\hline \hline
$x_{\rm cut}$ &                                  
$\eta^{\rm p}_{\rm d}/\eta^{\rm p}_{\rm u}$  &
$\eta^{\Lambda}_{\rm d}/\eta^{\Lambda}_{\rm s}$ \\
\hline\hline
$x_p>0.2$ & 0.637$\pm$0.173$\pm$0.083 & 0.468$\pm$0.069$\pm$0.030 \\ \hline
$x_p>0.3$ & 1.54 $\pm$1.18 $\pm$0.558 & 0.307$\pm$0.112$\pm$0.046 \\ \hline
$x_p>0.4$ & 0.335$\pm$0.234$\pm$0.109 & 0.118$\pm$0.141$\pm$0.076 \\ \hline
$x_p>0.5$ & 0.165$\pm$0.421$\pm$0.257 & 0.221$\pm$0.264$\pm$0.173 \\ \hline
\hline
\end{tabular}
\caption{Results for different values of the minimum $x_p$ cut
for $\rm uu$ diquark and strange diquark suppression as estimated by
$\eta^{\rm p}_{\rm d}/\eta^{\rm p}_{\rm u}$ and
$\eta^{\Lambda}_{\rm d}/\eta^{\Lambda}_{\rm s}$,
respectively, with statistical and systematic errors calculated taking 
into account correlations between the numerators and denominators 
of the ratios.   
\label{tab:baryonsupp}}
\end{center}
\end{table}

\begin{table}[p]
\begin{center}
\begin{tabular}{||l||c|c||}
\hline \hline
$x_{\rm cut}$ &                                  
$\eta^{\rm p}_{\rm u}/\eta^{\pi}_{\rm u}$  &
$\eta^{\Lambda}_{\rm s}/\eta^{\rm K}_{\rm s}$ \\
\hline\hline
$x_p>0.2$ & 0.149$\pm$0.023$\pm$0.017 & 0.184$\pm$0.030$\pm$0.007 \\ \hline
$x_p>0.3$ & 0.096$\pm$0.056$\pm$0.030 & 0.162$\pm$0.041$\pm$0.006 \\ \hline
$x_p>0.4$ & 0.188$\pm$0.052$\pm$0.016 & 0.170$\pm$0.071$\pm$0.014 \\ \hline
$x_p>0.5$ & 0.125$\pm$0.078$\pm$0.030 & 0.200$\pm$0.117$\pm$0.013 \\ \hline
\hline
\end{tabular}
\caption{Results for different values of the minimum $x_p$ cut
for $\rm ud$ diquark suppression as estimated by
$\eta^{\rm p}_{\rm u}/\eta^{\pi}_{\rm u}$ and
$\eta^{\Lambda}_{\rm s}/\eta^{\rm K}_{\rm s}$,
with statistical and systematic errors calculated taking 
into account correlations between the numerators and denominators 
of the ratios.   
\label{tab:baryonmeson}}
\end{center}
\end{table}


\newpage

\newpage

\newpage

\newpage

\newpage

\clearpage
\newpage

\end{document}